\documentclass[12pt]{article}
\usepackage[dvips]{graphicx}
\usepackage{amsfonts}
\usepackage{amssymb}
\usepackage{amsmath}
\usepackage{epsfig}
\usepackage{cite}
\usepackage{multirow}
\usepackage{latexsym}
\usepackage{graphicx}
\usepackage{float}

\newcommand{\nn}{\nonumber}
\newcommand{\be}{\begin{equation}}
\newcommand{\ee}{\end{equation}}
\newcommand{\bea}{\begin{eqnarray}}
\newcommand{\eea}{\end{eqnarray}}

\newcommand{\half}{\frac{1}{2}}

\newcommand{\beqa}{\begin{eqnarray}}
\newcommand{\eeqa}{\end{eqnarray}}

\newcommand{\Z}{\mathbb{Z}}

\newcommand{\un}{{\bf 1}}

\newcommand{\f}{{\bf 5}}
\newcommand{\fb}{{\bf \bar{5}}}
\newcommand{\te}{{\bf 10}}
\newcommand{\teb}{{\bf \bar{10}}}
\newcommand{\op}{\oplus}

\newcommand{\ad}{\mathrm{ad}}

\topmargin
-1.5cm
\textwidth
15.5cm
\textheight
23.5cm
\oddsidemargin
0.7cm
\evensidemargin
1.2cm

\begin{document}

\makeatletter
\@addtoreset{equation}{section}
\makeatother
\renewcommand{\theequation}{\thesection.\arabic{equation}}
\pagestyle{empty}
\rightline{CPHT-RR126.1209}
\vspace{1.9cm}
\begin{center}
\Large{\bf Froggatt-Nielsen models from $E_8$ in F-theory GUTs  \\[12mm]}
\large{Emilian Dudas$^{1,2}$,\; Eran Palti$^1$ \\[5mm]}
\small{$^1$ Centre de Physique Th´eorique, Ecole Polytechnique, CNRS, 91128 Palaiseau, France.}\\
\small{$^2$ LPT, UMR du CNRS 8627, Bˆat 210, Universit´e de Paris-Sud, 91405 Orsay Cedex, France.} \\[3mm]
\small{E-mail: emilian.dudas@cpht.polytechnique.fr, palti@cpht.polytechnique.fr} \\[12mm]
\small{\bf Abstract} \\[3mm]
\end{center}
\begin{center}
\begin{minipage}[h]{16.0cm}
This paper studies F-theory $SU(5)$ GUT models where the three generations of the standard model come from three different curves. All the matter is taken to come from curves intersecting at a point of enhanced $E_8$ gauge symmetry. Giving a vev to some of the GUT singlets naturally implements a Froggatt-Nielsen approach to flavour structure. A scan is performed over all possible models and the results are filtered using phenomenological constraints. We find a unique model that fits observations of quark and lepton masses and mixing well. This model suffers from two drawbacks: R-parity must be imposed by hand and there is a doublet-triplet splitting problem.
\end{minipage}
\end{center}
\newpage
\setcounter{page}{1}
\pagestyle{plain}
\renewcommand{\thefootnote}{\arabic{footnote}}
\setcounter{footnote}{0}


\tableofcontents


\section{Introduction}

The Standard Model (SM) of particle physics (and its extensions) relies on a substantial number of input parameters. The parameters, such as the quark masses, exhibit some basic patterns. Explaining these structures is the motivation for much of the proposed extensions of the SM. 
The structures are ideal for studying within string theory phenomenology since they do not rely on exact numbers but on symmetries and order-of-magnitude expressions. The study of flavour structures in Yukawa couplings of Grand Unified Theories (GUTs) has recently gained new impetus within string phenomenology with the developments of effective theories, labeled F-theory GUTs, where the top quark Yukawa coupling is of order one \cite{Donagi:2008ca,Beasley:2008dc,Beasley:2008kw,Donagi:2008kj}. This is certainly a promising starting point for a theory of flavour. Apart from this promising result these models are attractive because of their calculability properties: so called local models can decouple the bulk manifold for the purpose of many calculations. In this paper we will restrict to local model building, which means we decouple the bulk manifold. This type of model building forms a large body of literature, see \cite{Aldazabal:2000sa} for some of the original work and \cite{08052943,Hayashi:2008ba,Beasley:2008kw,Donagi:2008kj,Heckman:2008qt,Font:2008id,Marsano:2008jq,Heckman:2008qa,Blumenhagen:2008aw,09013785,Hayashi:2009ge,Bouchard:2009bu,09033009, Donagi:2009ra,Randall:2009dw,09043101,09043932,09052289,
Marsano:2009gv,Heckman:2009mn,09063297,Conlon:2009kt,Conlon:2009qa,Font:2009gq,08070789,Cecotti:2009zf,Conlon:2009qq,Hayashi:2009bt,Marchesano:2009rz,Chung:2009ib} for more recent developments within F-theory. Of course a fully global model is required for a consistent string vacuum and we refer to \cite{Blumenhagen:2008zz,09024143,09060013,Blumenhagen:2009yv,Marsano:2009gv,Donagi:2009ra,09043932,Marsano:2009wr} for recent developments in this direction. In particular \cite{Beasley:2008kw,Font:2008id,Heckman:2008qa,Bouchard:2009bu,Randall:2009dw,09043101,Heckman:2009mn,Font:2009gq,Cecotti:2009zf,Conlon:2009qq,Hayashi:2009bt,Marchesano:2009rz} studied the issue of flavour structures in F-theory models.

Local F-theory GUTs can be succinctly summarised as follows. The SM is realised on a 7-brane wrapping a contractible 4-dimensional manifold $S$ and carrying a GUT gauge group (which in this paper is $SU(5)$). This brane intersects other 7-branes along curves on its world-volume and along these curves there is localised matter which transforms in the bi-fundamental representation of the GUT group and an extra $U(1)$ associated to the intersecting brane. This matter can be thought of as coming from the adjoint representation of an enhanced gauge group along the matter curve. Interactions between different matter curves occur locally where they intersect and the gauge group is further enhanced. So for example Yukawa couplings are associated to the intersection of 3 matter curves where the gauge group is enhanced from the GUT group by at least rank 2. An important property of these interactions is that they must respect not only the GUT gauge symmetries but also the symmetries of the $U(1)$s. The main idea of this paper is to use these extra symmetries to generate the observed structures in the SM interactions.  

Generically different interaction/intersection points will arise at different locations on $S$. However in special circumstances it may be that there is a large enhancement at a certain point where many curves intersect and in particular it may be that the point enhanced gauge group is $E_8$. This point of $E_8$ is particularly interesting because enough curves intersect to give rise to all the interactions of the SM \cite{Heckman:2009mn,Marsano:2009gv}. From a calculability perspective this is very attractive since essentially all the possible interactions are simply governed by the local enhanced symmetry and can be studied, to a first approximation, irrespective of the distribution of curves on the whole of $S$. The more global aspects are encoded in the local monodromies of the curves about the intersection point \cite{Hayashi:2009ge,Donagi:2009ra,Heckman:2009mn,Marsano:2009gv}. A further attractive quality of models based on a point of $E_8$ is that they are `protected' against extensions, by which we mean that no more matter curves can intersect the point without inducing exotic states such as tensionless strings \cite{Heckman:2009mn,Katz:1996xe,Bershadsky:1996nu,Witten:1995gx}. Overall these type of models are ideal for studying symmetries and patterns within the SM interactions. 

In \cite{Heckman:2009mn,Marsano:2009gv} studies of such models were initiated. With regards to flavour structures and hierarchies the setup was such that the 3 generations all arose from a single matter curve: 1 curve for the 3 generations of $\te$s and 1 for the $\fb$s. The hierarchical structure was then invoked through the mechanism proposed in \cite{Heckman:2008qa} where the rank 1 Yukawas are corrected at subleading orders in an expansion parameter by effects induced through brane world-volume flux. Further studies of this mechanism found that although gauge flux alone does not induce such corrections \cite{Cecotti:2009zf,Conlon:2009qq}, the proposed corrections can arise through non-commutative deformations \cite{Cecotti:2009zf,Marchesano:2009rz}. In this paper we adopt a different approach to the flavour structure problem. We take the generations to arise from distinct matter curves. So 3 curves for the $\te$s and 3 for the $\fb$s. This means that each generation has different quantum numbers under the enhanced gauge symmetry. Such a setup naturally lends itself to applications of the Froggatt-Nielsen mechanism \cite{Froggatt:1978nt}. 

In the paper we study such embeddings of matter curves and Froggatt-Nielsen models within a point of $E_8$. We show that simple phenomenological constraints are enough to narrow thousands of possibilities to a handful of models. Imposing the crude phenomenological requirements of rank 1 renormalisable Yukawa couplings, rank 3 Yukawa couplings with Froggatt-Nielsen fields, no observable proton decay, and three light neutrinos rules out all the possible models. However if R-parity is imposed by hand\footnote{More precisely we require that terms of the type $\beta_i \f_{H_u} \fb_{i}$ with $i$ a family index are forbidden.} then we find that a number of models are possible and further imposing constraints on quark and neutrino masses and mixing we find a unique model that satisfies all the phenomenological constraints.\footnote{If we allow the Froggatt-Nielsen fields to come in vector-like pairs then a few more models are phenomenologically allowed. The problem with allowing vector-like pairs is that we expect them to gain a large mass and decouple from the low-energy effective theory. Due to this issue these models are relegated to the appendix.} 

Apart from having to impose R-parity by hand the model also suffers from a doublet-triplet splitting problem. The mechanism of hypercharge flux splitting \cite{Beasley:2008kw,Donagi:2008kj} is not available for these models. Indeed we show that this is the case for all local 3-curve models, i.e. models where the 3 generations come from 3 different curves. 

The paper is structured as follows. In section \ref{sec:matcurfn} we briefly review some details regarding F-theory GUTs and then study general features of Froggatt-Nielsen models within this framework. In section \ref{sec:phenoconst3} we impose phenomenological constraints on models based on 3 different curves for the generations. We find that we have to impose R-parity which leads to a unique model. In section \ref{sec:candmod} we further study this model and show that it matches phenomenological observations to a decent level. In section \ref{sec:furconst} we derive a no-go result for doublet-triplet splitting in 3-curve models. We summarise the paper in section \ref{sec:summary}. In appendix A we present models that satisfy crude phenomenological constraints but fail finer ones. In appendix B we present models where some of the Froggatt-Nielsen fields form vector pairs. In appendix \ref{sec:susy} we present a brief analysis of the vacuum for the phenomenologically successful model.

\section{Matter curves, $E_8$, and Froggatt-Nielsen}
\label{sec:matcurfn}

We begin this section with a review of local constructions of F-theory GUTs. The aim is to reach the model building rules that are used in the rest of the paper. In section \ref{sec:locmodf} we review local models in F-theory and in section \ref{sec:rolee8} we review the role that $E_8$ plays in local model building. The reviews are based on the papers \cite{Donagi:2008ca,Beasley:2008dc,Hayashi:2009ge,Donagi:2009ra,Heckman:2009mn,Marsano:2009gv}. In section \ref{sec:fnfguts} we study the application of the Froggatt-Nielsen mechanism.

\subsection{Local models of F-theory GUTs}
\label{sec:locmodf}

Our starting point is considering an F-theory 7-brane with gauge group $G_S=SU(5)$ wrapping a two complex-dimensional submanifold $S$ inside an elliptically fibered CY four-fold $X$. $S$ is usually taken to be contractible to a point in order for there to exist a limit where the bulk decouples. There will be other 7-branes inside $X$ wrapping submanifolds $S_i$ with gauge groups $G_i=U(1)_i$. These intersect the gauge brane S on matter (one complex-dimensional) curves $\Sigma_i = S \cap S_i$ along which two chiral multiplets localise which are charged under $G_S\times G_i$. The matter curves are associated to an enhanced (from $G_S$) gauge group $G_{\Sigma_i}\supset G_S \times G_i$ which is enhanced generically by rank 1. Matter curves associated to $\bf{5}$s of $SU(5)$ are associated with an enhanced gauge group $SU(6)$, while curves with $\bf{10}$ are associated to an enhancement to $SO(10)$. Turning on flux along the curve valued in the extra $U(1)$ generates chirality. 

When curves intersect at a point there is a Yukawa coupling induced at the intersection point, and there is a further enhancement which is generically by another rank 1. The intersections of the curves can be seen by decomposing the adjoint of the enhanced gauge group under $G_S\times G_i$. So that for example for the $\bf{66}$ of $SO(12)$ and the $\bf{78}$ of $E_6$ we have
\bea
SO(12) &\supset& SU(5) \times U(1)_a \times U(1)_b \;, \\
\bf{66} &\rightarrow& \bf{24}^{(0,0)} \op \un^{(0,0)} \op \un^{(0,0)} \op \left(\f \op \fb\right)^{(-1,0)} \op \left(\f \op \fb\right)^{(1,1)} \op
\left(\te \op \teb\right)^{(0,1)} \;, \nn \\
E_6 &\supset& SU(5) \times U(1)_{a'} \times U(1)_{b'} \;, \\
\bf{78} &\rightarrow& \bf{24}^{(0,0)} \op \un^{(0,0)} \op \un^{(0,0)} \op \un^{(-5,-3)} \op \un^{(5,3)} \nn \\
 & & \op \left(\f \op \fb\right)^{(-3,3)} \op \left(\te \op \teb\right)^{(-1,-3)} \op \left(\te \op \teb\right)^{(4,0)} \;. \nn
\eea
where superscripts denote $U(1)$ charges (of the $\f$ and $\te$, with $\fb$ and $\teb$ have opposite charges). For example we see that the point of $E_6$ comes from an intersection of three curves with $\f\op\fb$, $\te\op\teb$, and $\te\op\teb$. The Yukawa couplings, in this case the up-type, come from the interaction $\left(\ad_{G_P}\right)^3$. So the generic F-theory GUT has the Yukawa coupling associated to points of $E_6$ and $SO(12)$ enhancement.

\subsection{The role of $E_8$}
\label{sec:rolee8}

Local models in F-theory can also be described in a slightly different way. We can consider a 7-brane wrapping $S$ but now with a gauge group of higher rank than $SU(5)$. We then turn on a vacuum expectation value (vev) for the Higgs field in the world-volume theory such that at a generic point on $S$ the gauge group is broken to $SU(5)\times U(1)^4$. The vev vanishes along matter curves where the gauge group is therefore locally enhanced. This amends itself well to the viewing the 7-brane as a singularity in $X$ which is then resolved by the Higgs vev. Then all such models can be viewed locally as a resolution of an $E_8$ singularity, i.e. as a fibration of $S$ over an ALE space with an $E_8$ singularity. The generic deformation of an $E_8$ singularity to $SU(5)$ can be written as 
\be
y^2 = x^3 + b_5xy + b_4x^2z + b_3yz^2 + b_2xz^3 + b_0z^5 \;. \label{e8tosu5sing}
\ee
Here $y$, $x$ and $z$ are complex coordinates that are perpendicular to $S$ within $X$ such that $S$ is located at $z=0$. The $b_i$ vary over $S$ and so encode the fibration. 

The geometry is encoded in the decomposition of the adjoint of the enhanced gauge group
\bea
E_8 &\supset& SU(5) \times SU(5)_{\perp} : \nonumber \\
\bf{248} &\rightarrow& \left(\bf{24},\bf{1}\right)\op\left(\bf{1},\bf{24}\right)\op\left(\te,\f\right)\op\left(\fb,\te\right)\op\left(\teb,\fb\right)\op\left(\f,\teb\right)
\label{ade8}
\eea
This gives the number of curves, which all have to meet for the full $E_8$ to be restored. The connection to (\ref{e8tosu5sing}) is through the fact that the $b_i$ are given by elementary symmetric polynomials of the weights $t_i$, $i=1,..,5$ of the $\f$ of $SU(5)_{\perp}$. More concretely we associate a weight $t_i$ to a two-cycle that is blown up in resolving $E_8 \rightarrow SU(5)$ and each of the curves in (\ref{ade8}) can be written in terms of these as
\bea
\Sigma_{\te\op\teb} &:& t_i = 0 \;, \label{cur1} \\
\Sigma_{\f\op\fb} &:& -t_i - t_j = 0, \;\; i\neq j\;, \label{cur2} \\
\Sigma_{\bf{1}} &:& \pm \left(t_i - t_j\right) = 0 , \;\; i\neq j\;. \label{cur3}
\eea
We see that these are indeed 5 $\te\op\teb$ curves, 10 $\f\op\fb$ curves, and $2\times 10 + 4 =24$ singlets (here we get 2 singlets on each of the 10 matter curves in (\ref{cur3}) and there are an extra 4 that are not localised on a curve). In terms of the field theory the Higgs vev is in the Cartan subgroup of $SU(5)_{\perp}$ and so is given by 4 co-ordinates (the vev along the 4 Cartan generators of $SU(5)_{\perp}$) $\alpha_I$ with $I=1,2,3,4$. These correspond to the four-roots of E8 which are connected to the weights of the $\f$ (see \cite{Donagi:2008ca} for explicit relations). Since there are 4 roots and 5 weights there is one linear relation satisfied by the weights which is 
\be
t_1+t_2+t_3+t_4+t_5=0 \;. \label{sumconst}
\ee

The Higgs breaks $SU(5)_{\perp}$ to 4 $U(1)$s. These $U(1)$s are the $U(1)$s that the chirality generating flux is turned on and so generally they will all be anomalous and/or gain a Green-Schwarz mass. In the effective theory below the string scale they therefore may act as global symmetries. The charge of the matter curves under these are given in terms of the weights $t_i$ and since there are 4 linearly independent $U(1)$s and 4 linearly independent $t_i$s the requirement for a gauge invariant operator ${\cal O}$ involving fields from the matter curves is that when the curves are replaced by their root representatives as in (\ref{cur1}-\ref{cur3}) and then summed over they sum to zero. This can either occur through `mesonic' operators such as $\f \fb$ or through `baryonic' operators such as $\fb \fb \te$.

In specifying a local model the $b_i$ in (\ref{e8tosu5sing}) must be specified as functions of the coordinates on $S$. Although the $b_i$ are functions of the $t_i$ this does not specify the $t_i$ uniquely since the relations are non-linear. Solving for the $t_i$ one encounters branch cuts which correspond to monodromies relating the different $t_i$. The monodromy groups are subgroups of the Weyl group of $SU(5)$ and it was shown in \cite{Hayashi:2009ge} that wavefunctions of localised matter that are interchanged by monodromy actions are connected such that they should both be regarded as giving rise to the same matter zero mode. A more technologically advanced description of this phenomenon is made through the introduction of the spectral cover \cite{Donagi:2009ra}. We return to this description in section \ref{sec:furconst} but for now it is not required. The monodromy group is essential in understanding matter interactions. Each matter curve in (\ref{cur1}-\ref{cur3}) will lie in an orbit of the monodromy group and each element of this orbit represents the same field. Therefore interactions involving this field can arise from any gauge invariant combinations made by any of the representatives in the orbit. 

In this paper we study models that are based on a single point of $E_8$ enhanced gauge symmetry. The reason for this is that we would like to study matter interactions using only the quantum numbers and not their geometric properties. Since a point of $E_8$ is where all the curves in (\ref{cur1}-\ref{cur3}) geometrically meet, if an interaction between those fields is allowed by symmetries we expect, and therefore assume, that it will generically be present with an order 1 coefficient. If the curves were spread throughout $S$ there are possible geometric factors to do with potential separation of curves that must be taken into account and can modify the order 1 coefficients. Concentrating on a single point has benefits and drawbacks. These are the typical properties of models associated with a more bottom-up perspective than a top-down one: the most important drawbacks are of course that the models lack a global completion, which could be much more constrained, and therefore do not form true solutions to string theory. The main benefits are that there is much more freedom in model building. In particular for our case we allow ourselves the freedom to choose the chirality of fields on the matter curves. This corresponds to the fact that the chirality depends on integrating the flux over the whole curve and so relies on `global' properties of $S$ which a model based on a single point can decouple from. Nevertheless there are some constraints that still apply which we return to in section \ref{sec:furconst}.

To summarise, the model building setup we use is as follows. We consider a single point of $E_8$ which has the matter curves (\ref{cur1}-\ref{cur3}) intersecting. We are free to choose the net chirality of each curve. We are also free to choose the monodromy group that acts on the curves. With these specified we study the possible interactions with the rule that if an interaction is allowed by the symmetries of the problem then it is present with an order 1 coefficient. Finally we also allow ourselves to give a non-vanishing vacuum expectation value to up to 4 of the GUT singlets in (\ref{cur3}). These vevs will play a crucial role in our analysis and we now turn to a more detailed description of this aspect.

\subsection{Froggatt-Nielsen and F-theory GUTs}
\label{sec:fnfguts}

In this section we discuss applications of the Froggatt-Nielsen (FN) mechanism \cite{Froggatt:1978nt} (see \cite{Dreiner:2003hw,Babu:2009fd} for reviews) to the models described in the previous section. The basic idea is to generate flavour structures from vevs of GUT singlets localised on curves. In this study we allow ourselves to choose the vevs for the GUT singlets. However there is some motivation for why these singlets should develop a vev. The basic idea comes from the fact that since there is non-trivial flux associated to the 4 $U(1)$s that are in $SU(5)_{\perp}$ they may, and we assume do, become massive through the Green-Schwarz mechanism. This in turn means that they should induce D-terms with geometric moduli dependent Fayet-Iliopoulos (FI) terms. These D-terms will involve the open-string matter fields that are charged under the $U(1)$s which include the singlets. Now if the geometric moduli take values such that their contribution to the D-term is non-vanishing the matter fields will develop a vev to minimise the D-term. Since the GUT group should not be broken in this way the primary candidates are the singlets. Of course a proper treatment of the vacuum is really a global issue and in a local approach the singlet vevs can not be completely determined. We present a brief analysis of the `local' vacuum structure of a model in appendix \ref{sec:susy} simply to show the type of issues that arise. 

The question of whether the $U(1)$s gain a mass is a global one. Recall that if a $U(1)$ is anomalous it is guaranteed to gain a Green-Schwarz mass but it also gains such a mass if it couples to a globally non-trivial Ramond-Ramond field. Calculating this coupling within the local framework of this paper is not possible and therefore we assume that the $U(1)$s gain a mass as they should phenomenologically.\footnote{The coupling can be determined for the case where both the 2-cycle and dual 4-cycle associated to the $U(1)$ are local. In orbifold language these are $N=1$ sectors. Indeed the mass of such $U(1)$s can be determined by studying the mixed anomalies. In the case of a GUT there is only the $SU(5)$ mixed anomaly which implies one $U(1)$ must become massive.} 

With the $U(1)$s being massive they act effectively as global symmetries of the low energy effective theory. However because some of the GUT singlets develop a vev they break these symmetries spontaneously. If the singlets develop a large mass then we can write an effective theory where the singlets are integrated out. In this theory the global symmetries will therefore only be approximate. For a symmetry to remain a global symmetry to low energies the GUT singlets that develop a vev must not be charged under the associated $U(1)$.

The singlets that gain a vev should preferably be chiral, i.e. not form vector-like pairs. This is achieved by turning on flux along the singlet curves. The reason is that we expect a vector-like pair to gain a string-scale mass and so for them to develop a vev requires an FI larger than the string scale. Nonetheless such a setup is not completely excluded and we do present studies of models with vector-like pairs of singlets but due to these unattractive properties relegate them to the appendix.

The singlet vevs appear in the low energy effective theory divided by a suppression scale. The resulting small numbers, which we henceforth denote as $\epsilon_i$s are the expansion parameters for the FN mechanism. For much of the calculation the absolute value of the suppression scale is not important since it is only its ratio to the vevs, which we choose by hand, that affects the low-energy physics. However the suppression scale does appear explicitly in some other parts of the calculation such as the neutrino masses. We denote the suppressions scale $M_*$. In this paper we work in the context of local models and in that case the candidate suppression scale is the GUT scale
\be
M_* \simeq 2 \times 10^{16} \mathrm{\;GeV \;.}
\ee
Here we use the results of \cite{Conlon:2007zza} which show that the correct suppression scale should be the `winding' scale $M_s {\cal V}^{1/6}$, where ${\cal V}$ is the $B_6$ or CY 3-fold volume, and $M_s$ is the string scale.\footnote{The argument of \cite{Conlon:2007zza} applies to non-renormalisable operators in the superpotential, it may be that the supression scale for non-renormalisable operators coming from the Kahler potential differs from this scale.} This is combined with the results of \cite{Conlon:2009qa,Conlon:2009xf,Conlon:2009kt} which show that this is also the unification scale for a local model. 

\section{Phenomenological constraints on 3-curve models}
\label{sec:phenoconst3}

In this section we impose phenomenological constraints on all the possible Froggatt-Nielsen models arising from a point of $E_8$. We begin be studying constraints that can be imposed on just the matter curves. In section \ref{sec:mongrp} we constrain the possible monodromy groups and in section \ref{sec:rank1dy} we impose that the quark Yukawa matrices should be at most rank 1 at the renormalisable level, i.e. with no insertions of the singlet vevs. In section \ref{sec:barlepvio} we impose the constraints coming from baryon and lepton number violating operators. In section \ref{sec:muterm} we discuss the generation of a $\mu$ term through the Giudice-Masiero mechanism. We then go on to study the constraints on the models including the singlet vevs. Since there are a large number of such models the classification is performed using a computer program. In section \ref{sec:phenomix} we summarise the phenomenological constraints coming from quark and neutrino masses and mixings. Then in section \ref{sec:searchres} we report on the results of the computer search. 

\subsection{Monodromy groups}
\label{sec:mongrp}

Since we require the generations to arise from distinct curves we have the following 5 possibilities for the monodromy groups.\footnote{Note that we require at least a $\Z_2$ monodromy to make a perturbative $\f \te \te$ coupling.} 

\subsubsection*{Case 1: $G=\Z_2$}

This is based on a $\Z_2$ monodromy group that interchanges $t_1 \leftrightarrow t_2$. We have the $\te$ curves
\be
\te_1 : \left\{t_1,t_2\right\} \;,\;\; \te_2 : t_3 \;,\;\; \te_3 : t_4 \;,\;\; \te_4 : t_5 \;, \label{tecurz2}
\ee
where the curly brackets denote the orbit of the curve. Here, and henceforth, the matter curve $\te_1$ is associated to the top quark generation. We impose already that the top quark mass matrix should have exactly rank 1 without any singlet insertions. The other curves are left undetermined with 2 out of the 3 corresponding to the matter curves. The $\f$ and singlet matter curves are 
\bea
\f_{H_u} : -t_1 - t_2 \;,\;\; & & \un_1 : \pm\{t_1-t_3,t_2-t_3\} \;, \nn \\
\f_1 : \{-t_1-t_3,-t_2-t_3\} \;,\;\; & & \un_2 : \pm\{t_1-t_4,t_2-t_4\}  \;, \nn \\
\f_2 : \{-t_1-t_4,-t_2-t_4\} \;,\;\; & & \un_3 : \pm\{t_1-t_5,t_2-t_5\}  \;, \nn \\
\f_3 : \{-t_1-t_5,-t_2-t_5\} \;,\;\; & & \un_4 : \pm(t_3-t_4) \;, \nn \\
\f_4 : -t_3-t_4 \;,\;\; & & \un_5 : \pm(t_3-t_5) \;, \nn \\
\f_5 : -t_3-t_5 \;,\;\; & & \un_6 : \pm(t_4-t_5) \;, \nn \\
\f_6 : -t_4-t_5 \;,\;\; & & \un_7 : \{t_1-t_2,t_2-t_1\} \;.
\eea
Here we have identified the curve corresponding to the up Higgs. The other $\f$ matter curves comprise the Higgs down and 3 matter generations.\footnote{The singlet $\un_7$ is neutral under the remaining $U(1)$s and shares an orbit with its conjugate representation. We note that giving a string-scale vev to this field has similar effects to the monodromy group.} 

The remaining $\te$ and $\f$ curves correspond to exotic fields. We decouple them by not turning on flux on them so that they form vector pairs and gain a GUT-scale mass.

\subsubsection*{Case 2: $G=\Z_2\times \Z_2$}

This is based on a $\Z_2\times \Z_2$ monodromy group that interchanges $t_1 \leftrightarrow t_2$ or $t_3 \leftrightarrow t_5$. We have the $\te$ curves
\be
\te_1 : \left\{t_1,t_2\right\} \;,\;\; \te_2 : \left\{t_3,t_5\right\} \;,\;\; \te_3 : t_4 \;.
\ee
The other matter curves are
\bea
\f_{H_u} : -t_1 - t_2 \;,\;\; & &  \un_1 : \pm\{t_1-t_3,t_2-t_3,t_1-t_5,t_2-t_5\} \;, \nn \\
\f_1 : \{-t_1-t_3,-t_2-t_3,-t_1-t_5,-t_2-t_5\} \;,\;\; & &  \un_2 : \pm\{t_1-t_4,t_2-t_4\} \;, \nn \\
\f_2 : \{-t_1-t_4,-t_2-t_4\} \;,\;\; & &  \un_3 : \pm\{t_3-t_4,t_5-t_4\} \;, \nn \\
\f_3 : \{-t_3-t_4,-t_5-t_4\} \;,\;\; & &  \un_4 : \{t_3-t_5,t_5-t_3\} \;, \nn \\
\f_4 : -t_3-t_5 \;,\;\; & & \un_5 : \{t_1-t_2,t_2-t_1\} \;.
\eea

\subsubsection*{Case 3: $G=\Z_2$}

This is based on a $\Z_2$ monodromy group that interchanges $t_1 \leftrightarrow t_2$ and $t_3 \leftrightarrow t_5$ at the same time. We have the $\te$ curves
\be
\te_1 : \left\{t_1,t_2\right\} \;,\;\; \te_2 : \left\{t_3,t_5\right\} \;,\;\; \te_3 : t_4 \;.
\ee
The other matter curves are
\bea
\f_{H_u} : -t_1 - t_2 \;,\;\; & &  \un_1 : \pm\{t_1-t_3,t_2-t_5\} \;, \nn \\
\f_1 : \{-t_1-t_3,-t_2-t_5\} \;,\;\; & &  \un_2 : \pm\{t_1-t_4,t_2-t_4\} \;, \nn \\
\f_2 : \{-t_1-t_4,-t_2-t_4\} \;,\;\; & &  \un_3 : \pm\{t_1-t_5,t_2-t_3\} \;, \nn \\
\f_3 : \{-t_1-t_5,-t_2-t_3\} \;,\;\; & &  \un_4 : \pm\{t_3-t_4,t_5-t_4\} \;, \nn \\
\f_4 : \{-t_3-t_4,-t_5-t_4\} \;,\;\; & &  \un_5 : \{t_3-t_5,t_5-t_3\} \;, \nn \\
\f_5 : -t_3-t_5 \;,\;\; & & \un_6 : \{t_1-t_2,t_2-t_1\} \;.
\eea

\subsubsection*{Case 4: $G=\Z_3$}

This is based on a $\Z_3$ monodromy group that interchanges acts as $t_1 \rightarrow t_2 \rightarrow t_3 \rightarrow t_1$. We have the $\te$ curves
\be
\te_1 : \left\{t_1,t_2,t_3\right\} \;,\;\; \te_2 : t_4 \;,\;\; \te_3 : t_5 \;.
\ee
The remaining $\f$ and singlet curves are
\bea
\f_{H_u} : \{-t_1 - t_2,-t_1-t_3,-t_2-t_3\} \;, \;\; & & \un_1 : \pm\{t_1-t_4,t_2-t_4,t_3-t_4\} \;, \nn \\
\f_1 : \{-t_1-t_4,-t_2-t_4,-t_3-t_4\} \;, \;\; & & \un_2 : \pm\{t_1-t_5,t_2-t_5,t_3-t_5\} \;, \nn \\
\f_2 : \{-t_1-t_5,-t_2-t_5,-t_3-t_5\} \;, \;\; & & \un_3 : \pm(t_4-t_5) \;, \nn \\
\f_3 : -t_4-t_5 \;, \;\; & & \un_4 : \pm\{t_1-t_2,t_2-t_3,t_3-t_1\} \;.
\eea

\subsubsection*{Case 5: $G=S_3$}

This is based on a $S_3$ monodromy group that acts with permutations on $(t_1t_2t_3)$. We have the $\te$ curves
\be
\te_1 : \left\{t_1,t_2,t_3\right\} \;,\;\; \te_2 : t_4 \;,\;\; \te_3 : t_5 \;.
\ee
The remaining $\f$ and singlet curves are
\bea
\f_{H_u} : \{-t_1 - t_2,-t_1-t_3,-t_2-t_3\} \;, \;\; & & \un_1 : \pm\{t_1-t_4,t_2-t_4,t_3-t_4\} \;, \\
\f_1 : \{-t_1-t_4,-t_2-t_4,-t_3-t_4\} \;, \;\; & & \un_2 : \pm\{t_1-t_5,t_2-t_5,t_3-t_5\} \;, \nn \\
\f_2 : \{-t_1-t_5,-t_2-t_5,-t_3-t_5\} \;, \;\; & & \un_3 : \pm(t_4-t_5) \;, \nn \\
\f_3 : -t_4-t_5 \;, \;\; & & \un_4 : \{t_1-t_2,t_1-t_3,t_2-t_3,t_2-t_1,t_3-t_1,t_3-t_2\} \;. \nn
\eea

This concludes the classification of possible monodromy groups compatible with 3 matter curves. 

\subsection{Rank 1 down Yukawas}
\label{sec:rank1dy}

We impose on our models that the rank of the renormalisable down-type Yukawa matrix should be no larger than 1. This is a simple phenomenological requirement to avoid 2 heavy generations. For the different monodromy cases we have the choices of possible curves for the down Higgs and for generations of the $\te$s and $\f$s. We also impose that for a rank 1 down-type Yukawa, the quark that obtains a large mass is the bottom quark. This amounts to requiring that the non-vanishing entry is the one with the same $\te$ matter curve as that for the up-type Yukawas. 

Note that it is possible for the Yukawa matrices to be rank 0 at the renormalisable level and then generate the Yukawas with singlet insertions. For the up-type Yukawas this is not such an attractive possibility because of the large top mass but for the down-type Yukawas it is slightly preferred since for not too large tan $\beta$ the bottom Yukawa is suppressed compared to the top one. 

In this section we systematically go through the possibilities. 

\subsubsection*{Ruling out Cases 4 and 5}

The monodromy cases 4 and 5 act in the same way on the $\f$s and $\te$s. Therefore for the purposes of this section they can be treated on an equal footing. The main problem with these cases is that there are very few $\f$ matter curves. Indeed we are forced to choose the $\f_{H_d}$ to be localised on the same curve as the $\f_{H_u}$. This implies that a bare $\mu$ term is induced. Apart from this the renormalisable down-type Yukawas are
\bea
\fb_{H_d} \fb_1 \te_3 \;, \nn \\
\fb_{H_d} \fb_2 \te_2 \;, \nn \\
\fb_{H_d} \fb_3 \te_1 \;.
\eea
This means that the down-type Yukawa matrix is rank 3. We therefore rule out these monodromy possibilities since they do not lead to hierarchical quark masses.

\subsubsection*{Cases 1, 2 and 3}

Since there are more curves we can choose which curves have $H_d$. For case 1 we also need to choose the $\te$ curves while for cases 2 and 3 they are fixed. In case 1 we can take the matter curves to be $\te_1 \te_2 \te_3$. The cases $\te_1 \te_2 \te_4$ and $\te_1 \te_3 \te_4$ are generated by the reparameterisation symmetries $t_3 \leftrightarrow t_5$ and $t_4 \leftrightarrow t_5$. So this choice fixes these symmetries but note that we are still left with $t_3 \leftrightarrow t_4$ which we can use to relate $\f_1 \leftrightarrow \f_2$ and $\f_5 \leftrightarrow \f_6$. These are the reparameterisation symmetries we use for case 1. 

\begin{table}
\center
\begin{tabular}{|c|c|c|c|c|c|}
\hline
Case & $\fb_{H_d}$ & Allowed down-type Yukawas & Rank 0 & Rank 1 & b-quark\\
\hline
1 & $\fb_{H_u}$ & $\fb_{H_d} \fb_4 \te_4 \;,\; \fb_{H_d} \fb_5 \te_3 \;,\; \fb_{H_d} \fb_6 \te_2$ & $\checkmark$ & $\checkmark$ & $\times$ \\
\hline
1 & $\fb_{1}$ & $\fb_{H_d} \fb_2 \te_4 \;,\; \fb_{H_d} \fb_3 \te_3 \;,\; \fb_{H_d} \fb_6 \te_1$ & $\checkmark$ & $\checkmark$ & $\checkmark$\\  
\hline
1 & $\fb_{3}$ & $\fb_{H_d} \fb_1 \te_3 \;,\; \fb_{H_d} \fb_2 \te_2 \;,\; \fb_{H_d} \fb_4 \te_1$ & $\times$ & $\checkmark$ & $\checkmark$\\  
\hline
1 & $\fb_{4}$ & $\fb_{H_d} \fb_3 \te_1$ & $\checkmark$ & $\checkmark$ & $\checkmark$\\  
\hline
1 & $\fb_{5}$ & $\fb_{H_d} \fb_2 \te_1$ & $\checkmark$ & $\checkmark$ & $\checkmark$\\  
\hline
2 & $\fb_{H_u}$ & $\fb_{H_d} \fb_3 \te_2 \;,\; \fb_{H_d} \fb_4 \te_3$ & $\times$ & $\checkmark$ & $\times$\\  
\hline
2 & $\fb_{1}$ & $\fb_{H_d} \fb_2 \te_2 \;,\; \fb_{H_d} \fb_3 \te_1$ & $\times$ & $\times$ & $\checkmark$\\  
\hline
2 & $\fb_{2}$ & $\fb_{H_d} \fb_4 \te_1 \;,\; \fb_{H_d} \fb_1 \te_2$ & $\times$ & $\times$ & $\checkmark$\\  
\hline
2 & $\fb_{3}$ & $\fb_{H_d} \fb_1 \te_1$ & $\times$ & $\checkmark$ & $\checkmark$\\  
\hline
2 & $\fb_{4}$ & $\fb_{H_d} \fb_2 \te_1$ & $\times$ & $\checkmark$ & $\checkmark$\\  
\hline
3 & $\fb_{H_u}$ & $\fb_{H_d} \fb_{4} \te_2 \;,\; \fb_{H_d} \fb_5 \te_3$ & $\checkmark$ & $\checkmark$ & $\times$\\  
\hline
3 & $\fb_{1}$ & $\fb_{H_d} \fb_{2} \te_2 \;,\; \fb_{H_d} \fb_{4} \te_1$ & $\times$ & $\checkmark$ & $\checkmark$\\  
\hline
3 & $\fb_{2}$ & $\fb_{H_d} \fb_{1} \te_2 \;,\; \fb_{H_d} \fb_3 \te_2 \;,\; \fb_{H_d} \fb_5 \te_1$ & $\times$ & $\times$ & $\checkmark$\\  
\hline
3 & $\fb_{3}$ & $\fb_{H_d} \fb_{2} \te_2 \;,\; \fb_{H_d} \fb_{4} \te_1$ & $\times$ & $\checkmark$ & $\checkmark$\\  
\hline
3 & $\fb_{4}$ & $\fb_{H_d} \fb_1 \te_1 \;,\; \fb_{H_d} \fb_3 \te_1$ & $\times$ & $\checkmark$ & $\checkmark$\\  
\hline
3 & $\fb_{5}$ & $\fb_{H_d} \fb_2 \te_1$ & $\times$ & $\checkmark$ & $\checkmark$\\  
\hline
\end{tabular}
\label{case1downrank}
\caption{Table showing the allowed Yukawa couplings for different monodromy cases with various choices for the Higgs down curve. The last 3 columns show if the down-type Yukawa matrix can be of rank 0 and rank 1 for an appropriate choice of $\f$ matter curves and if the massive down-type quark is the bottom quark.}
\end{table}
In table 1 we present the results. The two columns before the last show if for some choice of matter curves it is possible to make the down-type Yukawa at most rank 0 and rank 1 respectively. Models that do not satisfy this constraint are ruled out. The last column shows if the quark that gains a mass is the bottom quark. Three of the models do not satisfy this. The same models also have the up and down Higgs coming from the same curve and therefore a bare $\mu$ term is not forbidden. We phenomenologically rule out these models.

For each choice of $\f_{H_d}$ that survives the constraints imposed there is still choice for the $\f$ matter curves. The only possibilities compatible with at most rank 1 bottom-type Yukawa matrix are listed in table 2. There the models are denoted according to the notation `$\mathrm{\bf Mono\;case}.\fb_{H_d}.\fb_{M1}.\fb_{M2}.\fb_{M3}$'. For example $1.1.2.6.4$ corresponds to monodromy case 1, the Higgs down curve is $\fb_1$ and the families are $\fb_2$, $\fb_6$ and $\fb_4$. We now go on to further restrict these models using phenomenological constraints.

\subsection{Constraints from Baryon and Lepton number violation}
\label{sec:barlepvio}

There are strong phenomenological constraints coming from baryon and lepton number violating operators in the presence of TeV scale supersymmetry. For a review see \cite{Barbier:2004ez}. The important operators are 
\bea
W &\supset& \beta_i \fb_M^i \f_{H_u} + \lambda_{ijk} \fb_M^i \fb_M^j \te_M^k + W^1_{ijkl} \te_M^i \te_M^j \te_M^k \fb_M^l \nn \\
 & &+ W^2_{ijk}  \te_M^i \te_M^j \te_M^k \fb_{H_d} + W^3_{ij} \fb_M^i \fb_M^j \f_{H_u}\f_{H_u} + W^4_{i} \fb_M^i \fb_{H_d} \f_{H_u} \f_{H_u} \;, \nn \\
 K &\supset& K^1_{ijk} \te_M^i \te_M^j \f^k_{M} + K^{2}_i \fb_{H_u} \fb_{H_d} \te_M^i \;.
\eea
Here $W$ denotes terms coming from the superpotential and $K$ from the Kahler potential. Note that $\beta$, $\lambda$, $W_2$, $W_4$, $W_5$, $K_1$ and $K_2$ are all R-parity violating. 

Proton decay forms one of the strongest constraints. It constrains $W^1$ by itself and $W^2$ and $K^1$ are severely constrained as products with $\lambda$. For an order of magnitude estimate we have that \cite{Barbier:2004ez,Smirnov:1996bg}
\bea
\lambda_{ijk} &\leq& 10^{-5} \;\forall\;\{i,j,k\} \;\;, \;\lambda_{111} \leq 10^{-12}\;, \nn \\  
W^1_{112l} &\leq& 10^{-7} \frac{M_{*}}{M_p} \sim 10^{-10}\;.
\eea
Here the indices denote generations with label 1 the lightest. The limit on $W^1$ is given for particular generations. However it also limits heavier generations through quark mixing. Taking the mixing to be compatible with the CKM we find that at the renormalisable level, i.e. with no singlet insertions, any generation combination is ruled out.\footnote{The case where the three quarks are all of the heaviest generation is marginal but anyway such an operator is not compatible with the $U(1)$ gauge symmetries.} 

The operator $W^3$ induces neutrino masses but as long as the suppression scale is larger than $10^{13}\mathrm{\;GeV}$ this is acceptable.

The $\beta$ term is very strongly constrained. In an exact supersymmetric theory, in the absence of further quantum numbers distinguishing the down Higgs and the lepton doublets it can be rotated away. However such a general rotation is not always possible in the F-theory GUT case because there are global symmetries, inherited from the massive $U(1)$s, that distinguish between the down Higgs and the leptons. Although some of these symmetries may be broken by the singlet vevs some of them may survive to low energies. Anyway such a rotation can not get rid of both the superpotential $\beta$ terms and the corresponding soft terms simultaneously unless they are nearly exactly correlated. Even then trilinear R-parity violating terms $\lambda$ are induced which must be suppressed so that $\beta$ should be at most of order the $\mu$-term scale \cite{Barbier:2004ez}. The presence of the $\beta$ terms, even if they are suppressed to the $\mu$-term scale, generates large masses for the neutrinos. Indeed they should be suppressed at around $\beta \leq 10^{-22} M_*$ to maintain lighter than eV neutrino masses.  

If all the curves meet at a single point of $E_8$ we expect that any gauge invariant operators should be of order 1 up to the mass dimension suppression by the GUT scale. If the curves are spread throughout $S$ then there may be some geometric suppression factors. Since we are looking at the former possibility we should certainly forbid the possible constrained combinations. In table 2 we display the presence or absence of each term at the renormalisable level, i.e. with no insertions of the singlets. 
\begin{table}
\center
\begin{tabular}{|c|c|c|c|c|c|c|c|c|}
\hline
Model & $\beta$ & $\lambda$ & $W^1$ & $W^2$ & $W^3$ & $W^4$ & $K^1$ & $K^2$ \\
\hline
\hline
\multicolumn{9}{|c|}{Rank 0 models} \\
\hline
1.1.2.4.5  &  & $\checkmark$  &   &  &    &  & $\checkmark$ &  \\
\hline
1.4.1.2.5  &  & $\checkmark$  &   &  &    &  & $\checkmark$ &  \\
\hline
1.4.1.2.6  &  & $\checkmark$  &   &  &    &  & $\checkmark$ &  \\
\hline
1.4.1.5.6  &  &  $\checkmark$ &   &  &    &  & $\checkmark$ &  \\
\hline
1.4.2.5.6  &  & $\checkmark$  &   &  &    &  & $\checkmark$ &  \\
\hline
1.5.1.3.4  &  & $\checkmark$  & $\checkmark$  &  &    &  & $\checkmark$ &  \\
\hline
1.5.1.3.6  &  &  $\checkmark$ & $\checkmark$  &  &    &  & $\checkmark$ &  \\
\hline
1.5.1.4.6  &  &  $\checkmark$ &   &  &  &    & $\checkmark$ &  \\
\hline
1.5.3.4.6  &  & $\checkmark$  & $\checkmark$  &  &    &  & $\checkmark$ &  \\
\hline
\hline
\multicolumn{9}{|c|}{Rank 1 models} \\
\hline
1.1.6.2.4  &  &   &   &  &  &    & $\checkmark$ &  \\
\hline
1.1.6.2.5  &  & $\checkmark$    &   &  &  &  & $\checkmark$ &  \\
\hline
1.1.6.4.5  &  &   &   &   &    &  & $\checkmark$ &  \\
\hline
1.3.4.5.6  &  &   &   & $\checkmark$ &  &    & $\checkmark$ &  \\
\hline
1.4.3.1.2  &  & $\checkmark$ & $\checkmark$ &   &  &    & $\checkmark$ &  \\
\hline
1.4.3.1.5  &  & $\checkmark$ & $\checkmark$ &   &  &    & $\checkmark$ &  \\
\hline
1.4.3.1.6  &  & $\checkmark$ & $\checkmark$ &   &  &    & $\checkmark$ &  \\
\hline
1.4.3.5.6  &  &   & $\checkmark$ &   &  &    & $\checkmark$  &  \\
\hline
1.5.2.1.3  &  & $\checkmark$ & $\checkmark$ &   &  &  &   $\checkmark$ &  \\
\hline
1.5.2.1.4  &  &   &   &   &  &  &   $\checkmark$ &  \\
\hline
1.5.2.1.6  &  & $\checkmark$ &   &   &  &  &   $\checkmark$ &  \\
\hline
1.5.2.3.4  &  & $\checkmark$ & $\checkmark$ &   &  &  &   $\checkmark$ &  \\
\hline
1.5.2.3.6  &  & $\checkmark$ & $\checkmark$ &   &  &  &   $\checkmark$ &  \\
\hline
1.5.2.4.6  &  &   &   &   &  &  &   $\checkmark$ &  \\
\hline
2.3.1.2.4  &  & $\checkmark$ & $\checkmark$ &   &  &    & $\checkmark$ &  \\
\hline
2.4.2.1.3  &  & $\checkmark$ & $\checkmark$ &   &  &    & $\checkmark$ &  \\
\hline
3.1.4.3.5  &  & $\checkmark$ & $\checkmark$ & $\checkmark$ &  &   & $\checkmark$ &  \\
\hline
3.3.4.1.5  &  & $\checkmark$ & $\checkmark$ & $\checkmark$ &  &   & $\checkmark$ &  \\
\hline
3.4.1.2.5  &  & $\checkmark$ & $\checkmark$ & $\checkmark$ &  &    & $\checkmark$ &  \\
\hline
3.4.3.2.5  &  & $\checkmark$ & $\checkmark$ & $\checkmark$ &  &    & $\checkmark$ &  \\
\hline
3.5.2.1.3  &  & $\checkmark$ & $\checkmark$ & $\checkmark$ &  &    & $\checkmark$ &  \\
\hline
3.5.2.1.4  &  & $\checkmark$ & $\checkmark$ & $\checkmark$ &  &    & $\checkmark$ &  \\
\hline
3.5.2.3.4  &  & $\checkmark$ & $\checkmark$ & $\checkmark$ &  &    & $\checkmark$ &  \\
\hline
\end{tabular}
\label{protondecayconst}
\caption{Table showing the presence or absence (by $\checkmark$ and blank space respectively) of lepton and baryon-number violating operators. The models are denoted by the notation `$\mathrm{\bf Mono\;case}.\fb_{H_d}.\fb_{M1}.\fb_{M2}.\fb_{M3}$'}
\end{table}
Only 5 models survive the constraints with no extra symmetries and a further 8, denoted in brackets, are allowed if we impose R-parity by hand
\bea
1.1.6.2.4 \;&,& \;\; (1.1.2.4.5) \nn \\
1.1.6.4.5 \;&,& \;\; (1.4.1.2.5) \nn \\
1.3.4.5.6 \;&,& \;\; (1.4.1.2.6) \nn \\
1.5.2.1.4 \;&,& \;\; (1.4.1.5.6) \nn \\
1.5.2.4.6 \;&,& \;\; (1.4.2.5.6) \nn \\
(1.1.6.2.5)\;&,& \;\; (1.5.1.4.6) \nn \\
(1.5.2.1.6) \;&.& \label{goodmod}
\eea
Here the models are denoted by $\mathrm{\bf Mono\;case}.\fb_{H_d}.\fb_{M1}.\fb_{M2}.\fb_{M3}$.

Since we are also giving a vev to some of the singlets we should make sure that the operators are forbidden to the required order even after singlet insertions. For the case of the $\beta$ and $\mu$ terms this practically means forbidden at all orders.

\subsection{The $\mu$ term}
\label{sec:muterm}

Since we are forbidding a $\mu$ term in the superpotential we should generate it through the Giudice-Masiero mechanism \cite{Giudice:1988yz}. This amounts to requiring a coupling in the Kahler potential\footnote{It is interesting to note that if the renormalisable Yukawa couplings $\f_{H_u} \te_i \te_j$ and $\fb_{H_d} \te_k \fb_l$ are allowed by the symmetries then the proton decay operator $\te_i\te_j\te_k\fb_l$ has the opposite charge to $\f_{H_u}\fb_{H_d}$ so that a Giudice-Masiero coupling in the Kahler potential implies a coupling of the type $Y\te_i\te_j\te_k\fb_l$ in the superpotential. Since the Yukawas are only non-vanishing at the renormalisable level for the heaviest generations this only puts mild restrictions on the vev $y$.}
\be
Y^{\dagger} \f_{H_u} \fb_{H_d} \;, \label{guidice}
\ee
where $Y$ develops an F-term and also possibly a vev
\be
Y = y + \theta^2 F_Y \;.
\ee
The field $Y$ can be a GUT singlet that is from a curve intersecting the point of $E_8$, i.e. an $X$ field. However it could also be a GUT singlet that is localised on a curve which does not intersect the point of $E_8$. If this curve forms a triple intersection with the curves associated to $\f_{H_u}$ and $\fb_{H_d}$ the operator (\ref{guidice}) would be induced. This possibility is important for the model we present in section \ref{sec:candmod} where there are no candidate singlets to form the coupling (\ref{guidice}) that could develop an F-term. Note that in this setup the coupling of $Y$ to fields other than $\f_{H_u}$ and $\fb_{H_d}$ would be geometrically suppressed.

\subsection{Constraints from quark and neutrino masses and mixing}
\label{sec:phenomix}

Recreating the masses and mixing of the SM fields is the primary reason for adopting the Froggatt-Nielsen approach. In the quark sector the masses and mixing are determined by the Yukawa couplings. These essentially need to recreate a hierarchical structure in the quark masses and small mixing angles between the quarks. The masses should roughly fit the GUT values (see for example \cite{Babu:2009fd})\footnote{These are quoted for tan $\beta = 50$.}
\bea
m_u \sim 5 \times 10^{-7} \mathrm{\;TeV} \;, \;\; m_c \sim 2 \times 10^{-4} \mathrm{\;TeV} \;, \;\; m_t \sim 0.1 \mathrm{\;TeV} \;, \nn \\ 
m_d \sim 5 \times 10^{-7} \mathrm{\;TeV} \;, \;\; m_s \sim 10^{-5} \mathrm{\;TeV} \;, \;\; m_b \sim 6 \times 10^{-4} \mathrm{\;TeV} \;. \label{expquarkmass}
\eea
The mixing in the quark sector should match the Cabibbo-Kobayashi-Maskawa (CKM) matrix \cite{Amsler}
\be
\mathrm{V_{CKM}} = \left( \begin{array}{ccc} 
V_{ud} & V_{us} & V_{ub} \\ 
V_{cd} & V_{cs} & V_{cb}\\ 
V_{td} & V_{ts} & V_{tb}
\end{array} \right)
\sim \left( \begin{array}{ccc} 
1 & 0.2 & 0.004 \\ 
0.2 & 1 & 0.04 \\ 
0.009 & 0.04 & 1
\end{array} \right)
\ee

For a review of the relevant neutrino physics see \cite{Strumia:2006db}. The neutrino sector data is given by the mass splitting \cite{Amsler}
\be
\Delta m^2_{12} \sim 10^{-4} \mathrm{\;eV}^2 \;,\;\;\Delta m^2_{23} \sim 10^{-3} \mathrm{\;eV}^2 \;.
\ee
The mixing in the neutrino sector is given by the Pontecorvo-Maki-Nakagawa-Sakata (PMNS) matrix which for our level of accuracy reads\footnote{Note that our conventions for diplaying the matrix differ from the standard literature. The standard notation can be reached from (\ref{genpmns}) by transposing and interchanging the labels for the right-handed neutrino index $1\leftrightarrow 3$.}
\be
\mathrm{U_{PMNS}} = \left( \begin{array}{ccc} 
U_{e 3} & U_{\mu 3} & U_{\tau 3} \\ 
U_{e 2} & U_{\mu 2} & U_{\tau 2}\\ 
U_{e 1} & U_{\mu 1} & U_{\tau 1}
\end{array} \right)
\sim \left( \begin{array}{ccc} 
1 & 1 & 1 \\ 
1 & 1 & 1 \\ 
(<0.2) & 1 & 1
\end{array} \right) \label{genpmns}
\ee
We note that atmospheric neutrino oscillations require $\nu_{\mu}\rightarrow\nu_{\tau}$ while solar neutrino oscillations require $\nu_{e}\rightarrow\nu_{\mu}$ or/and $\nu_{e}\rightarrow\nu_{\tau}$.

The model building in the neutrino sector is done by specifying 3 different curves each of which carries a right-handed neutrino. 
We then use two mechanisms to give masses to the neutrinos. The first is the traditional seesaw mechanism and the second is the Dirac mass scenario proposed in \cite{ArkaniHamed:2000bq,Bouchard:2009bu}. The relevant mass contributions come from the terms
\bea
& &\int d^2\theta \left( D_W H_u L  N + M N N \right) + \int d^4\theta \frac{D_K \left(H_d\right)^{\dagger} L \; N}{M_{*}} \nn \\
&=& \int d^2\theta \left( v D_W L  N + M N N + \frac{\mu}{M_{*}} v D_K L N\right) \;.
\eea
Here $H_u$ and $H_d$ denote the up and down Higgs superfields , $L$ denote the left-handed leptons, and $N$ denote the right-handed neutrino superfields. Following the equality we replaced $H_d$ by its F-term $F_{H_d}=\mu H_{u}$ assuming that a $\mu$ term is generated via the Giudice-Masiero mechanism, and replaced $H_u$ by its vev $<H_u>=v$. The relevant mass matrices are therefore the superpotential Dirac mass matrix $D_W$, the Majorana mass matrix $M$ and the Kahler potential Dirac mass matrix $D_K$. 

\subsection{Results of models scan}
\label{sec:searchres}

The scan over all the possible choices for matter and singlet curves was performed using a computer program. In this section we report on the results. 

A particular model is built as follows. We begin from a specification of the $\te$ and $\f$ matter curves that satisfy the constraints of the previous sections. A list of these is given in (\ref{goodmod}). For each of these choices we scan over the options of turning on a vev for some (up to 4) of the remaining singlet curves. For each such model we calculate the resulting interactions, including singlet insertions, as allowed by the quantum numbers. We then eliminate models according to phenomenological constraints on the quark sector. For each of the models that survive we scan over the possible choices of 3 singlet curves on which the right-handed neutrinos are localised. 

We impose the following constraints
\begin{enumerate}
\item The up and down quark Yukawa matrices are rank 1 without any singlet insertions.
\item The up and down quark Yukawa matrices are rank 3 with singlet insertions.
\item The $\mu$ term is forbidden with any number of singlet insertions.
\item The operator $W_1$ is forbidden if it only involves the 2 lighter generations and no more than 3 singlet insertions.
\item There are no vector pairs of fields.
\end{enumerate}
Condition 4 is imposed from proton decay constraints, see section \ref{sec:barlepvio}. Condition 5 is imposed since, according to our rules of allowing all operators allowed by symmetries, vector pairs of fields gain a UV scale mass. See section \ref{sec:fnfguts} for more regarding this issue. However for the case of the Froggatt-Nielsen singlets, since we take the vevs as input parameters, this constraint is not completely compulsory. Therefore we do perform a scan over models where the Froggatt-Nielsen singlets can come in vector pairs but treat the results in appendix B.

We have not imposed that the operators $\beta$ and $\lambda$ should vanish with singlet insertions. As discussed in section \ref{sec:barlepvio} this is required. However we found that there are no models that satisfy the conditions above and have the $\beta$ and $\lambda$ terms vanishing. This is true for any number of singlets with a vev. Therefore we conclude that {\bf in the absence of any further symmetries or mechanisms to suppress terms there are no phenomenologically viable Froggatt-Nielsen models}. 

This forces to impose a symmetry and the ideal candidate is R-parity (matter-parity). This forbids the problematic terms $\beta$ and $\lambda$. Actually we shall see that the only viable models already have $\lambda$ vanishing and so really we only require that $\beta$ is suppressed. 

Once we impose R-parity we find some phenomenologically viable models for the quark sector. These are all shown in appendix A. There are only six such models, three 2-field models and three 3-field models\footnote{Here we denote a model where $n$ of the singlets have a vev an $n$-field model. Note also that there are four 3-field models in the appendix but one has an identical quark sector to the 2-field model.}, and in the appendix we show that four of them give too large quark mixing to be compatible with CKM constraints. This leaves only two viable models for the quark sector. 

Having fixed the quark sector we turn to the neutrino sector. We take the RHN neutrinos to come from singlet matter curves. For more details see section \ref{sec:phenomix}. For the model scan we impose the following constraint
\begin{itemize}
\item There are 3 light neutrinos or 2 light neutrinos and 1 massless one.
\end{itemize}
This leaves a small number of models that are shown in the appendix. We find that 1 of the 2 viable quark sectors is consequently ruled out for not having a viable neutrino sector. This leaves a unique quark sector. This quark scenario has 6 possible neutrino scenarios. 5 are 2-field models and there is 1 3-field model. However if also impose that
\begin{itemize}
\item There is some mixing between all 3 generations of neutrinos,
\end{itemize}
so that no neutrino flavour decouples from the other 2 as required by the combination of atmospheric and solar neutrino oscillations. We are left with only 3 possible scenarios. 2 of these are discussed in detail in appendix \ref{sec:2fns} where it is argued that they either have too small mixing or too light neutrino to match observations. However we note that they are not far off being phenomenologically relevant. Ruling them out leaves a unique candidate model which is shown in table \ref{tab:goodmodel}. 

We note that, as shown in appendix B, if we allow for the Froggatt-Nielsen fields to come in vector pairs then there are more candidate models and in particular a phenomenologically attractive model is studied in appendix \ref{sec:vecexa1}.

\begin{table}
\begin{minipage}{0.5\textwidth}
  \centering
\begin{tabular}{|ccc|}
\hline
Field & Curve & Charges/Orbit \\
\hline
\hline
\multicolumn{3}{|c|}{Chiral spectrum} \\
\hline
$\f_{H_u}$  & $\f_{H_u}$ & $-t_1-t_2$ \\ 
$\fb_{H_d}$ & $\fb_{5}$ & $t_3+t_5$ \\ 
$\te_{t}$   & $\te_{1}$ & $\left\{t_1,t_2\right\}$ \\
$\te_{c}$   & $\te_{3}$ & $t_4$ \\
$\te_{u}$   & $\te_{2}$ & $t_3$ \\
$\fb_{b}$   & $\fb_{2}$ & $\{t_1+t_4,t_2+t_4\}$ \\
$\fb_{s}$   & $\fb_{1}$ & $\{t_1+t_3,t_2+t_3\}$ \\
$\fb_{d}$   & $\fb_{4}$ & $t_3+t_4$ \\
$N_1$       & $\bar{1}_{6}$ & $-t_4+t_5$ \\
$N_2$       & $\bar{1}_{5}$ & $-t_3+t_5$ \\
$N_3$       & $\bar{1}_{3}$ & $\{-t_1+t_5,-t_2+t_5\}$ \\
$X_1$       & $\bar{1}_{4}$ & $-t_3+t_4$ \\
$X_2$       & $1_{2}$ & $\{t_1-t_4,t_2-t_4\}$ \\
$X_3$       & $\bar{1}_{1}$ & $\{-t_1+t_3,-t_2+t_3\}$ \\ 
\hline \hline
\multicolumn{3}{|c|}{Non-Chiral spectrum} \\
\hline
- & $\te_{4}$ & $t_5$ \\
- & $\f_3$ & $\{-t_1-t_5,-t_2-t_5\}$ \\
- & $\f_6$ & $-t_4-t_5$ \\
- & $\un_7$ & $t_1-t_2$ \\
\hline
\end{tabular}
\end{minipage}
\begin{minipage}{0.5\textwidth}
\center
\begin{tabular}{|c|c|}
\hline
\multicolumn{2}{|c|}{Chiral interactions} \\
\hline \hline
$\f_{H_u} \te_i \te_j$ & $\left( \begin{array}{ccc} \epsilon_{2}^2\epsilon_{\bar{4}}^2 & \epsilon_{2}^2\epsilon_{\bar{4}} & \epsilon_{2}\epsilon_{\bar{4}} \\ \epsilon_{2}^2\epsilon_{\bar{4}} & \epsilon_{2}^2 & \epsilon_{2} \\ \epsilon_{2}\epsilon_{\bar{4}} & \epsilon_{2} & 1\end{array} \right) $ \\
\hline \hline
$\fb_{H_d}\fb_i \te_j$ & $\left( \begin{array}{ccc} \epsilon_{2}^2\epsilon_{\bar{4}}^2 & \epsilon_{2}\epsilon_{\bar{4}}^2 & \epsilon_{2}\epsilon_{\bar{4}} \\ \epsilon_{2}^2\epsilon_{\bar{4}} & \epsilon_{2}\epsilon_{\bar{4}} & \epsilon_{2} \\ \epsilon_{2}\epsilon_{\bar{4}} & \epsilon_{\bar{4}} & 1\end{array} \right)$ \\
\hline \hline
$K\supset \f_{H_d} \fb_i N_j$ & $\left( \begin{array}{ccc} \epsilon_{2} & 1 & \epsilon_{\bar{1}}\epsilon_{2} \\ \epsilon_{\bar{1}}\epsilon_{2} & \epsilon_{\bar{1}} & \epsilon_{\bar{1}}\epsilon_{\bar{1}}\epsilon_{2} \\ 1 & \epsilon_{\bar{1}}\epsilon_{\bar{4}} & \epsilon_{\bar{1}} 
\end{array} \right)$\\
\hline \hline
$\beta \f_{H_u} \fb_i$ & $\left(\epsilon_{\bar{4}}\epsilon_2^2,\epsilon_{\bar{4}}\epsilon_2,\epsilon_2\right)$ \\
\hline \hline
$\f_{H_u} \fb_i N_j$ & $0$ \\
\hline \hline
$M N_i N_j$ & $0$ \\
\hline \hline
$\fb_i \fb_j \te_k$ & 0 \\
\hline \hline
$\te_i \te_j \te_k \fb_l$ & 0 \\
\hline \hline
$\mu \f_{H_u} \fb_{H_d}$ & 0 \\
\hline
\end{tabular}
\vspace{0.75in}
\end{minipage}
\caption{Tables showing the spectrum and mass operators of a model that satisfies the imposed phenomenological constraints. Here $\epsilon_{\bar{4}}=\frac{<X_1>}{M_*}$, $\epsilon_2=\frac{<X_2>}{M_*}$ and $\epsilon_{\bar{1}}=\frac{<X_3>}{M_*}$. The indices $(i,j)$ label columns and rows respectively and range over $\{u,c,t\}$, $\{d,s,b\}$ for the quarks and leptons and $\{3,2,1\}$ for the right-handed neutrinos (with 1 denoting the heaviest). The monodromy group is the $\Z_2$ of case 1. The vanishing operators can all be attributed to a factor of $t_5$. Note that we are forced to impose R-parity by hand to forbid a $\beta$-term. }
\label{tab:goodmodel}
\end{table}

\section{A candidate model}
\label{sec:candmod}

In this section we study the unique model that survives the phenomenological constraints imposed in the models scan. The model is shown in table \ref{tab:goodmodel}. We split the phenomenological analysis into the quark sector and the neutrino sector that are studied in sections \ref{sec:goodmodquark} and \ref{sec:goodmodneut} respectively. 

It is useful to note a symmetry of the model that can be used to easily see the vanishing of many operators in table \ref{tab:goodmodel}. This symmetry is associated to the weight $t_5$ and follows from the fact that only the right-handed neutrino curves and $H_d$ contain a factor of $t_5$. In particular the singlets that develop a vev do not break this global symmetry that is inherited from a massive $U(1)$ which means it remains a good symmetry to low energies.\footnote{Although 3 singlets develop a vev they only break 2 linearly independent $U(1)$ combinations.} This means, for example, that there is no $\mu$ term generated to any order since the operator $\mu H_u H_d$ contains a factor of $t_5$ which can not be canceled by any singlet insertions. Similar considerations forbid the $W^1$ and $\lambda$ terms. This symmetry means that, as claimed, we only need to forbid the $\beta$ term by hand rather than using the full power of R-parity.\footnote{Note that under this global symmetry the down Higgs and the leptons have different quantum numbers so that it is not possible to rotate them. In particular it means that it is not possible to rotate the $\beta$ term away.}

There is another point worth noting. From table \ref{tab:goodmodel} it is easy to see that it is not possible to generate a Giudice-Masiero operator $Y^{\dagger}\f_{H_u}\fb_{H_d}$ from any of the singlets. This implies that the relevant field $Y$ should come from a curve that does not intersect the point of $E_8$. Hence this whole sector is decoupled from our calculation. In appendix \ref{sec:2fns} we present 2 models that can avoid this issue but are phenomenologically less attractive as models of neutrino masses and mixing. In appendix \ref{sec:vecexa1} we present a model that avoid this issue and is phenomenologically attractive but contains a vector pair of Froggatt-Nielsen fields.

Finally we note that from the pure singlet superpotential interaction $X_1 X_2 X_3$ all the singlets gain a large mass and can effectively by integrated out at low energies. In appendix \ref{sec:susy} we present a brief vacuum analysis of this model. This serves to highlight some tension between the singlet vevs and supersymmetry. Of course the analysis is incomplete since the vacuum is really a global issue that does not decouple from moduli stabilisation. 

\subsection{The quark sector}
\label{sec:goodmodquark}

The quark sector Yukawa couplings depend on the vevs $\epsilon_2$ and $\epsilon_{\bar{4}}$ which should be chosen to match observations. Going from the Yukawa matrices to the observable quark masses and mixing is, for a generic Yukawa matrix, non-trivial. However for an approximately diagonal Yukawa matrix it is possible to use the formula for the mixing parameters in the CKM matrix \cite{Hall:1993ni} 
\be
s_{12} \simeq \frac{Y_{12}}{Y_{22}} \;,\; s_{13} \simeq \frac{Y_{13}}{Y_{33}} \;,\; s_{23} \simeq \frac{Y_{23}}{Y_{33}} \;,
\ee
where the corresponding angles in the CKM matrix are given by $s^{CKM}_{ij} \simeq s^D_{ij} - s^U_{ij}$.
As long as these are small the approximation is valid. We then choose the vevs to match the Wolfenstein parameterisation of the mixing angles 
\be
s_{12} = \lambda \;,\;\; s_{23} \simeq \lambda^2 \;,\;\; s_{13} \simeq \lambda^3 \;, 
\ee
where the Wolfenstein parameter is the Cabbibo angle $\lambda \simeq 0.2$.
These imply the vevs
\be
\epsilon_{2} = \lambda^2 \;, \nn \\
\epsilon_{\bar{4}} = \lambda \;.
\ee
The resulting Yukawa and CKM matrices read
\be
Y^U \simeq \left( \begin{array}{ccc} 
\lambda^6 & \lambda^5 & \lambda^3 \\ 
\lambda^5 & \lambda^4 & \lambda^2\\ 
\lambda^3 & \lambda^2 & 1
\end{array} \right) \;,\;\;
Y^D \simeq \left( \begin{array}{ccc} 
\lambda^6 & \lambda^4 & \lambda^3 \\ 
\lambda^5 & \lambda^3 & \lambda^2 \\ 
\lambda^3 & \lambda & 1 
\end{array} \right) \;,\;\;
V_{CKM} \simeq \left( \begin{array}{ccc} 
1 & \lambda & \lambda^3 \\ 
\lambda & 1 & \lambda^2 \\ 
\lambda^3 & \lambda^2 & 1 
\end{array} \right) 
\;. \label{1521442yukawas}
\ee
We go on to analyse the phenomenology of this setup. But first we note a point that arises again in the upcoming sections. The matrices (\ref{1521442yukawas}) all have order one factors in front of each element and we simply denote their scaling with the singlet vevs. These factors arise from the geometry of the different curves and for higher order operators from integrating out heavy states. In the absence of symmetries the factors of the different elements are unrelated. Now the form of (\ref{1521442yukawas}) is such that neglecting these order one factors the determinant of the matrix vanishes. However since there are no symmetries that constrain these factors we do not expect the determinant to vanish once they are included.\footnote{Candidate symmetries are the permutation symmetries acting on the $t_i$s. These however are broken by the choice of matter curves and singlets with vevs.}

The resulting CKM matrix matches observations very well since it recreates the Wolfenstein form exactly. Note that we have chosen 2 parameters and fit the 6 elements of the CKM.

Since the Yukawas are approximately diagonal the quark masses can be read off the diagonal and take the ratios
\bea
& &\mathrm{Up\;quarks}\;\;\lambda^6:\lambda^4:1 \;, \\
& &\mathrm{Down\;quarks}\;\;\lambda^6:\lambda^3:1 \;.
\eea
These clearly have the required hierarchical structure. If we take the prefactors to be exactly 1 the quark masses normalised to the top and bottom quarks are
\bea
m_u \sim 6 \times 10^{-6} \mathrm{\;TeV} \;, \;\; m_c \sim 2 \times 10^{-4} \mathrm{\;TeV} \;, \;\; m_t \sim 0.1 \mathrm{\;TeV} \;, \nn \\ 
m_d \sim 4 \times 10^{-8} \mathrm{\;TeV} \;, \;\; m_s \sim 5 \times 10^{-6} \mathrm{\;TeV} \;, \;\; m_b \sim 6 \times 10^{-4} \mathrm{\;TeV} \;.
\eea
The masses are fairly consistent with the experimental values quoted in (\ref{expquarkmass}). 
Note that our methodology of fixing the singlet vevs is not really a best fit analysis to the quark masses and mixing but for the needs of this paper it is sufficient.

\subsection{The neutrinos sector}
\label{sec:goodmodneut}

The lepton sector phenomenology is determined by the down-type Yukawas $Y^D$ and the Dirac masses from the Kahler potential $D_K$. In this model there are no superpotential Dirac masses and there are no Majorana masses. The neutrinos are Dirac. The vevs for $\epsilon_{2}$ and $\epsilon_{\bar{4}}$ are fixed from the quark sector. This leaves the vev $\epsilon_{\bar{1}}$ as a free parameter for the neutrino sector. The mass matrix in table \ref{tab:goodmodel} has two eigenvalues of order 1 and one eigenvalue of order $\lambda^2\epsilon^2_{\bar{1}}$. Taking the down-Higgs vev at $200$GeV and the $\mu$ term to be 1TeV the vev $\epsilon_{\bar{1}}\simeq 0.1$ gives the not unreasonable neutrino masses 
\bea
m_{\nu_1} &\simeq& m_{\nu_2} \simeq \; 10^{-2} \mathrm{\;eV}, \nn \\
m_{\nu_3} &\simeq& 4 \times 10^{-6} \mathrm{\;eV}\;.
\eea

Determining the PMNS matrix for the Yukawas is non-trivial since the Dirac mass matrix for the neutrinos is not approximately diagonal. Indeed it leads to an essentially anarchic PMNS matrix. For example choosing the order 1 coefficients randomly in the range 0.8-1.2 gives the PMNS matrix\footnote{The actual values for $\{11,12,13,21,22,23,31,32,33\}$ are $\{1.20,1.03,0.85,0.83,0.85,1.16,1.18,1.05,0.80\}$.}
\be
U_{PMNS} \simeq \left( \begin{array}{ccc} 
-0.05 & -0.49 & -0.91 \\ 
0.94 & 0.30 & -0.14 \\ 
-0.11 & 0.91 & -0.35 
\end{array} \right) 
\ee
The coefficients vary considerably upon varying the random numbers (typical range is between 0.01-1). However there is no special feature that suppresses the $U_{e3}$ term and, in the absence of further symmetries, its smallness is not explained.

In summary we find that this model forms a decent fit to all the relevant observations.

\section{No-go for doublet-triplet splitting and 3-curve models}
\label{sec:furconst}

The models we have studied are based on a single point of $E_8$ symmetry. In this section we discuss some results that apply once we consider the whole of $S_{GUT}$ but not the full CY 4-fold. In doing this we restrict to particular models, by imposing some properties on $S_{GUT}$, but are able to make statements about mechanisms that the point approach is not sensitive to. In particular we show that in the `semi-local' approach where we can decouple $S_{GUT}$ from the bulk {\bf the mechanism of double-triplet splitting on the Higgs curves by hypercharge flux \cite{Beasley:2008kw,Donagi:2008kj} can not be applied to models where the generations come from 3 different curves unless the monodromy group is trivial.} The result is an extension of that presented in \cite{Marsano:2009gv}. The basic overview of our result is that, for 3-curve models, it is not possible to have a net hypercharge flux on any $\f$ matter curves without the flux also restricting non-trivially to one of the SM $\te$ curves. Since the hypercharge flux breaks the GUT structure having a non-trivial restriction to a SM $\te$ matter curve is forbidden. In section \ref{sec:result} we derive the claimed result and in section \ref{sec:implications} we briefly discuss its implications for our models.

\subsection{The no-go result}
\label{sec:result}

In this section we use extensively the results and formalism of \cite{Donagi:2009ra,Marsano:2009gv}. We refer the reader to these references for the introduction of the relevant concepts and definitions.

We begin by studying how the hypercharge flux restricts to the $\te$ matter curves inside $S_{GUT}$. In order to account for the monodromy properties of the $\te$ matter curves we need to introduce the spectral cover. More precisely we need to introduce the spectral cover for the fundamental representations of $SU(5)_{\perp}$. The spectral cover is a hypersurface inside the projective 3-fold
\be
X = {\mathbb P}({\cal O}_{S_{GUT}}\oplus K_{S_{GUT}}) \;,
\ee
given by the constraint
\be
C_{10} = b_0 U^5 + b_2 V^2 U^3 + b_3 V^3 U^2 + b_4 V^4 U + b_5 V^5 = 0 \;. \label{spect10}
\ee
Here ${\cal O}_{S_{GUT}}$ and $K_{S_{GUT}}$ are the trivial and canonical bundle on $S_{GUT}$ respectively and $\{U,V\}$ are homogeneous complex coordinates on the ${\mathbb P}^1$ fibre in $X$. The $b_i$ are the same as in (\ref{e8tosu5sing}) and are given by the elementary symmetric polynomials of degree $i$ in the $t_i$. The idea is that locally we can set some affine parameter $s=U/V$ in which (\ref{spect10}) is a polynomial whose 5 roots are exactly the $t_i$. Indeed $s$ can be equated with the value of the Higgs field and overall (\ref{spect10}) forms a 5-fold cover of $S_{GUT}$. The monodromy of the Higgs or the $t_i$ is encoded in the global properties of (\ref{spect10}) and more specifically in how the polynomial decomposes into products. We can think of all the $\te$ curves as lifting to a single curve on the spectral cover which then decomposes into parts according to the decomposition of the spectral cover. Indeed this curve is determined by the equation $U=0$ which gives
\be
b_5 = t_1 t_2 t_3 t_4 t_5 = 0 \;,
\ee
which reproduces the equations for the five $\te$ matter curves. 

Now consider the case where the spectral cover decomposes into 4 separate pieces so that it takes the form
\be
C_{10} = \left(a_1 V^2 + a_2 V U + a_3 U^2 \right) \left(a_4 V + a_7 U\right) \left(a_5 V + a_8 U\right) \left(a_6 V + a_9 U\right) = 0 \;. \label{spect10z2}
\ee
Here the $a_I$ are some as yet undetermined coefficients. This decomposition corresponds to a ${\mathbb Z}_2$ monodromy group that by choice of parameterisation we shall take to act as $t_1 \leftrightarrow t_2$. So that the $\te$ curves $t_1$ and $t_2$ both lift to a curve on a single factor of the spectral cover given by the first brackets in (\ref{spect10z2}). Now we can write the $b_i$ in terms of the $a_I$ as
\bea
b_0 &=& a_{3789} \;, \nn \\
b_1 &=& a_{2789} + a_{3678} + a_{3579} + a_{3489} \;, \nn \\
b_2 &=& a_{1789} + a_{2678} + a_{2579} + a_{2489} + a_{3567} + a_{3468} + a_{3459} \;, \nn \\
b_3 &=& a_{3456} + a_{1678} + a_{1579} + a_{1489} + a_{2567} + a_{2468} + a_{2459} \;, \nn \\
b_4 &=& a_{2456} + a_{1567} + a_{1468} + a_{1459} \;, \nn \\
b_5 &=& a_{1456} \;. \label{aIsol}
\eea
Here we use the notation $a_{IJKL}=a_I a_J a_K a_L$. 

We are interested in determining the curves $a_I=0$ on $S_{GUT}$. This can be done as follows. The $b_i$ transform as sections of the bundle $\eta - i c_1$ \cite{Donagi:2009ra,Marsano:2009gv}. Here $c_1$ is the first Chern class of the tangent bundle of $S_{GUT}$ and $\eta = 6 c_1 - t$ with $-t$ being the first Chern class of the normal bundle to $S_{GUT}$. Using (\ref{aIsol}) this then implies that the $a_I$ transform as shown in table \ref{tab:aIsect}.
\begin{table}
\centering
\begin{tabular}{|c|c|}
\hline
Section & $c_1$(Bundle)\\
\hline
$a_1$ & $\eta - 2c_1 - \left(\chi_7 + \chi_8 + \chi_9\right)$ \\
\hline
$a_2$ & $\eta - c_1 - \left(\chi_7 + \chi_8 + \chi_9\right)$ \\
\hline
$a_3$ & $\eta - \left(\chi_7 + \chi_8 + \chi_9\right)$\\
\hline
$a_4$ & $-c_1 + \chi_7$ \\
\hline
$a_5$ & $-c_1 + \chi_8$\\
\hline
$a_6$ & $-c_1 + \chi_9$\\
\hline
$a_7$ & $\chi_7$ \\
\hline
$a_8$ & $\chi_8$\\
\hline
$a_9$ & $\chi_9$\\
\hline
\end{tabular}
\caption{Table showing the first Chern classes of the line bundles that the $a_I$ are sections of. The forms $\chi_{\{7,8,9\}}$ are unspecified.}
\label{tab:aIsect}
\end{table}

With these specifications we can study how the hypercharge flux restricts to the $\te$ matter curves. First we recall that, in order to avoid inducing a Green-Schwarz mass, the hypercharge flux is given by the first Chern class of a line bundle for which the dual 2-cycle on $S_{GUT}$ is in the kernel of the map $H^2(S) \rightarrow H^2(B_3)$ \cite{Buican:2006sn,Donagi:2008kj,Beasley:2008kw}. In other words the hypercharge flux is on a 2-cycle that is non-trivial on $S_{GUT}$ but trivial in the full CY. In particular this implies 
\be
F_Y \cdot c_1 = F_Y \cdot (-t) = 0 \;.
\ee
Here $F_Y$ denotes the hypercharge flux and $\cdot$ the intersection. This follows from the fact that $c_1$ and $-t$ both correspond to 2-cycles that are non-trivial in $B_3$ \cite{Donagi:2009ra,Marsano:2009gv}. Therefore the hypercharge flux restricts trivially to these curves. Indeed in table \ref{tab:aIsect} the only possible classes that the hypercharge could restrict non-trivially to are the $\chi_i$. 

Now consider the $\te$ matter curves. These are given by $b_5=a_1a_4a_5a_6=0$. In terms of (\ref{tecurz2}) the SM matter curves $\te_1$, $\te_2$ and $\te_3$ are given by $a_1=0$, $a_4=0$ and $a_5=0$ respectively. If we require that the hypercharge must restrict trivially to these then we have 
\be
F_Y \cdot (\chi_7 + \chi_8 + \chi_9) = 0 \;,\;\; F_Y \cdot \chi_7 = 0 \;,\;\; F_Y \cdot \chi_8 = 0 \;.
\ee
These imply that the hypercharge must also restrict trivially to $\te_4$. Therefore we conclude that for this case it is not possible for the hypercharge to restrict non-trivially to any $\te$ matter curves. Indeed for a 3-curve model only a trivial monodromy group would allow a non-trivial restriction of hypercharge flux to some $\te$ matter curves while restricting trivially to the SM $\te$ curves. Once this is established we can use the no-go theorem of \cite{Marsano:2009gv} which states that if there is trivial hypercharge restriction to all the $\te$ matter curves there can not be any non-trivial restriction to the $\f$ matter curves. This in turn implies that we can not use the hypercharge flux to split the doublets and the triplets of the Higgs curves.

\subsection{Implications for the models}
\label{sec:implications}

The fact that for any 3-curve model with non-trivial monodromy group the hypercharge flux can not be used for doublet-triplet splitting is disappointing. Within the `semi-local' framework with $S_{GUT}$ being a contractible cycle and decoupling from the bulk there are no particularly attractive alternatives to the hypercharge flux for the job of doublet-triplet splitting. In order to accommodate this we could try to relax some of the phenomenological constraints we have imposed on our models. One possibility, suggested in \cite{Marsano:2009gv}, is to take the Higgs up and Higgs down to come from the same curve. This of course induces a $\mu$-term problem which seems very difficult to avoid. In our setup there is also a further problem. Performing a computer search over such models we find that either the up-type Yukawa or the down-type Yukawa matrix is at most rank 1 with any number of singlet insertions. This can be understood from the fact that, when $\f_{H_u}=\f_{H_d}$ the gauge invariance of one type of Yukawa implies the other type is not gauge invariant.\footnote{This is certainly true at the renormalisable level with no singlet insertions. But it seems to still hold even with singlet insertions.} 

Another possibility is to take the monodromy group trivial. This would imply that the renormalisable top Yukawa coupling vanishes but we can hope to recover it with a singlet insertion. This is phenomenologically less attractive since it would be suppressed. Indeed it must be suppressed by at least 2 different singlets which is a large suppression.\footnote{This is because for a trivial monodromy group only off-diagonal terms can be induced at the renormalisable level in the up-type Yukawa matrix. Therefore we require that it is rank 0 otherwise we would get large quark mixing. This implies that the top Yukawa should be of the form $t_1 + t_2 - t_4 - t_5$ which requires at least 2 different singlet insertions to be gauge invariant.} Anyway, performing a computer scan we find that there are no rank 3 Yukawa matrices generated with such a mechanism. 

A different approach is to embed the point of $E_8$ inside some global model where $S_{GUT}$ is not decoupled from the bulk. In this case all the possibilities are open again to solve the doublet-triplet problem and to break the GUT group. We refer to \cite{Donagi:2008kj} for some suggestions. 

Finally we note that, as suggested in \cite{Beasley:2008kw}, it may be possible that the hypercharge vanishes when integrated over the $\f$ curve but still restricts to it non-trivially pointwise. This could lead to some mass splitting in analogue with the Aharonov-Bohm effect. 

\section{Summary}
\label{sec:summary}

In this paper we studied the possibility of generating realistic flavour structure by applying the Froggatt-Nielsen mechanism to F-theory $SU(5)$ GUT models based on a point of enhanced $E_8$ gauge symmetry. The 3 generations of the SM came from 3 different curves so that each generation had different quantum numbers under the enhanced gauge group which played the role of the Froggatt-Nielsen global symmetry group. The Froggatt-Nielsen fields were taken to come from singlet curves that intersect the $E_8$ point but are not charged under the GUT $SU(5)$. All the field interactions were taken to be `natural' in that if an interaction is allowed by gauge symmetries it appears in the Lagrangian with the appropriate suppression scale.

We scanned through all possible configurations of this setup and imposed phenomenological constraints on the resulting interactions. We found that only a handful of models satisfied even crude constraints. Imposing slightly finer constraints such as hierarchical quark masses with small mixing and 3 light neutrinos with large mixing left only a unique possibility. This model was found to fit observations quite well.\footnote{Allowing for vector-like pairs of Froggatt-Nielsen fields lead to further phenomenologically attractive models which are studied in appendix B.} 

We found that generically realistic models suffered from two serious problems. The first being operators of the form $\beta \f_{H_u} \fb_M$ not being suppressed. This could be solved by imposing R-parity. The second being the lack of a simple mechanism for doublet-triplet splitting. Indeed we showed that for any 3-curve model the doublet-triplet splitting by hypercharge flux mechanism of \cite{Beasley:2008kw,Donagi:2008kj} can not be used. 

\subsection*{Acknowledgments}

We thank S. Lavignac and Graham Ross for useful discussions and Joe Conlon, Joe Marsano and Sakura Schafer-Nameki for very helpful and very patient explanations of their work and for reading through some of this manuscript.

The work in this paper was supported in part by the European ERC Advanced Grant 226371 MassTeV, by the CNRS PICS no. 3059 and 4172, 
by the grants ANR-05-BLAN-0079-02, and the PITN contract PITN-GA-2009-237920.

\appendix

\section{Semi-viable models}
\label{sec:semiviab}

In this appendix we go through some of the models that do not fail the `first-order' phenomenological constraints but do fail more refined criteria.
We split the appendix according to the number of vevs that are turned on. We display all the models that satisfy the constraints 1-5 of section \ref{sec:searchres}. The models are all labeled according to the appropriate curves. A negative value for a curve number means the conjugate representation. We display the relevant matrices according to 
\be
Y = \left( \begin{array}{ccc} 5_3 N_3 & 5_2 N_3 & 5_1 N_3 \\ 5_3 N_2 & 5_2 N_2 & 5_1 N_2 \\ 5_3 N_1 & 5_2 N_1 & 5_1 N_1 
\end{array} \right) \;.
\ee
Here $N_i$ stand either for right-handed neutrinos or for $\te$ matter curves as appropriate. The appropriate vevs are denoted by $\epsilon_{i}$ where again the index is negative for the vev of the conjugate field (in table \ref{tab:goodmodel} this is denoted as $\epsilon_{\bar{i}}$).

We split the analysis into models where two or three of the singlets develop a vev. There are no 1-field or 4-field models that forbid $\mu$ term and $W_1$ term and have rank 3 up and down Yukawas.

\subsection{$2$-field models}
\label{sec:2fsemiviab}

\subsubsection{Quark sector}

In this section models are labeled as
\be
\mathrm{\bf Mono\;case}.\fb_{H_d}.\fb_{M1}.\fb_{M2}.\fb_{M3}.\bf{1}_{S1}.\bf{1}_{S2} \;.
\ee
The possible models are:

\scriptsize

\be
\mathrm{Model}\;1.5.2.1.4.-4.2 : 
Y^U = \left( \begin{array}{ccc} \epsilon_{2}\epsilon_{2} & \epsilon_{2}\epsilon_{2}\epsilon_{-4} & \epsilon_{2} \\ \epsilon_{2}\epsilon_{2}\epsilon_{-4} & \epsilon_{2}\epsilon_{2}\epsilon_{-4}\epsilon_{-4} & \epsilon_{2}\epsilon_{-4} \\ \epsilon_{2} & \epsilon_{2}\epsilon_{-4} & 1\end{array} \right) \;,\;\;
Y^D = \left( \begin{array}{ccc} \epsilon_{2}\epsilon_{2}\epsilon_{-4} & \epsilon_{2}\epsilon_{-4} & \epsilon_{2} \\ \epsilon_{2}\epsilon_{2}\epsilon_{-4}\epsilon_{-4} & \epsilon_{2}\epsilon_{-4}\epsilon_{-4} & \epsilon_{2}\epsilon_{-4} \\ \epsilon_{2}\epsilon_{-4} & \epsilon_{-4} & 1\end{array} \right) \;.
\ee

\be
\mathrm{Model}\;1.5.2.1.4.1.2 : 
Y^U = \left( \begin{array}{ccc} \epsilon_{2}\epsilon_{2} & \epsilon_{1}\epsilon_{2} & \epsilon_{2} \\ \epsilon_{1}\epsilon_{2} & \epsilon_{1}\epsilon_{1} & \epsilon_{1} \\ \epsilon_{2} & \epsilon_{1} & 1\end{array} \right) \;,\;\;
Y^D = \left( \begin{array}{ccc} \epsilon_{1}\epsilon_{2} & \epsilon_{1} & \epsilon_{2} \\ \epsilon_{1}\epsilon_{1} & 0 & \epsilon_{1} \\ \epsilon_{1} & 0 & 1\end{array} \right) \;.
\ee

\be
\mathrm{Model}\;1.5.2.1.4.1.4 : 
Y^U = \left( \begin{array}{ccc} \epsilon_{1}\epsilon_{1}\epsilon_{4}\epsilon_{4} & \epsilon_{1}\epsilon_{1}\epsilon_{4} & \epsilon_{1}\epsilon_{4} \\ \epsilon_{1}\epsilon_{1}\epsilon_{4} & \epsilon_{1}\epsilon_{1} & \epsilon_{1} \\ \epsilon_{1}\epsilon_{4} & \epsilon_{1} & 1\end{array} \right) \;,\;\;
Y^D = \left( \begin{array}{ccc} \epsilon_{1}\epsilon_{1}\epsilon_{4} & \epsilon_{1} & \epsilon_{1}\epsilon_{4} \\ \epsilon_{1}\epsilon_{1} & 0 & \epsilon_{1} \\ \epsilon_{1} & 0 & 1\end{array} \right) \;.
\ee

\normalsize 

We can rule out the last two models $1.5.2.1.4.1.2$ and $1.5.2.1.4.1.4$. First they do not give rise to a viable neutrino sector (see next section). Secondly they give rise to large quark mixing. This can be seen as follows. Consider $1.5.2.1.4.1.2$ for example. We interchange $\f_2 \leftrightarrow \f_3$. This puts the matrix in roughly diagonal form and so the resulting CKM can be read off using the techniques of \cite{Hall:1993ni}. Since the off-diagonal angles are small we recover a roughly diagonal CKM matrix. However since we interchanged the down and strange quarks the large mixing is now between the charm and down quarks and up and strange quarks. 

Therefore there is a unique 2-field model that can recreate realistic physics in the quark sector. The model still has freedom within the neutrino sector which we now turn to.

\subsubsection{Neutrino sector}
\label{sec:2fns}

In this section we label the models according to
\be
\mathrm{\bf Mono\;case}.\fb_{H_d}.\fb_{M1}.\fb_{M2}.\fb_{M3}.\bf{1}_{S1}.\bf{1}_{S2}.\bf{1}_{RHN1}.\bf{1}_{RHN2}.\bf{1}_{RHN3} \;.
\ee

\scriptsize

\bea
\mathrm{Model}\;& &1.5.2.1.4.2.-4.-6.-5.-3.\;\;\;\;K = \left( \begin{array}{ccc} \epsilon_{2} & 1 & 0 \\ 0 & 0 & 0 \\ 1 & 0 & 0\end{array} \right) \;,\\ 
W &=& \left( \begin{array}{ccc} 0 & 0 & 0 \\ 0 & 0 & 0 \\ 0 & 0 & 0\end{array} \right) \;,\;\;\;M = \left( \begin{array}{ccc} 0 & 0 & 0 \\ 0 & 0 & 0 \\ 0 & 0 & 0\end{array} \right) \;.
\eea

\bea
\mathrm{Model}\;& &1.5.2.1.4.2.-4.-6.-5.-1.\;\;\;\;K = \left( \begin{array}{ccc} 0 & 0 & 0 \\ 0 & 0 & 0 \\ 1 & 0 & 0\end{array} \right) \;,\\ 
W &=& \left( \begin{array}{ccc} 0 & \epsilon_{2}\epsilon_{2}\epsilon_{-4}\epsilon_{-4} & \epsilon_{2}\epsilon_{2}\epsilon_{-4} \\ 0 & 0 & 0 \\ 0 & 0 & 0\end{array} \right) \;,\;\;\;M = \left( \begin{array}{ccc} \epsilon_{2}\epsilon_{2}\epsilon_{-4}\epsilon_{-4} & 0 & 0 \\ 0 & 0 & 0 \\ 0 & 0 & 0\end{array} \right) \;.
\eea

\bea
\mathrm{Model}\;& &1.5.2.1.4.2.-4.-6.-3.-1.\;\;\;\;K = \left( \begin{array}{ccc} 0 & 0 & 0 \\ \epsilon_{2} & 1 & 0 \\ 1 & 0 & 0\end{array} \right) \;,\\ 
W &=& \left( \begin{array}{ccc} 0 & \epsilon_{2}\epsilon_{2}\epsilon_{-4}\epsilon_{-4} & \epsilon_{2}\epsilon_{2}\epsilon_{-4} \\ 0 & 0 & 0 \\ 0 & 0 & 0\end{array} \right) \;,\;\;\;M = \left( \begin{array}{ccc} \epsilon_{2}\epsilon_{2}\epsilon_{-4}\epsilon_{-4} & 0 & 0 \\ 0 & 0 & 0 \\ 0 & 0 & 0\end{array} \right) \;.
\eea

\bea
\mathrm{Model}\;& &1.5.2.1.4.2.-4.-6.-1.3.\;\;\;\;K = \left( \begin{array}{ccc} 0 & 0 & 0 \\ 0 & 0 & 0 \\ 1 & 0 & 0\end{array} \right) \;,\\ 
W &=& \left( \begin{array}{ccc} 0 & 0 & 0 \\ 0 & \epsilon_{2}\epsilon_{2}\epsilon_{-4}\epsilon_{-4} & \epsilon_{2}\epsilon_{2}\epsilon_{-4} \\ 0 & 0 & 0\end{array} \right) \;,\;\;\;M = \left( \begin{array}{ccc} 0 & 0 & 0 \\ 0 & \epsilon_{2}\epsilon_{2}\epsilon_{-4}\epsilon_{-4} & 0 \\ 0 & 0 & 0\end{array} \right) \;.
\eea

\bea
\mathrm{Model}\;& &1.5.2.1.4.2.-4.-5.-3.-1.\;\;\;\;K = \left( \begin{array}{ccc} 0 & 0 & 0 \\ \epsilon_{2} & 1 & 0 \\ 0 & 0 & 0\end{array} \right) \;,\\ 
W &=& \left( \begin{array}{ccc} 0 & \epsilon_{2}\epsilon_{2}\epsilon_{-4}\epsilon_{-4} & \epsilon_{2}\epsilon_{2}\epsilon_{-4} \\ 0 & 0 & 0 \\ 0 & 0 & 0\end{array} \right) \;,\;\;\;M = \left( \begin{array}{ccc} \epsilon_{2}\epsilon_{2}\epsilon_{-4}\epsilon_{-4} & 0 & 0 \\ 0 & 0 & 0 \\ 0 & 0 & 0\end{array} \right) \;.
\eea

\normalsize 

All these models can recreate semi-realistic masses for the neutrinos. However if we impose more refined constraints we can rule more of them out. 
In the model $1.5.2.1.4.2.-4.-6.-5.-3$ the $\nu_{\tau}$ completely decouples from the other 2 which is inconsistent with atmospheric neutrino oscillations. While in the models $1.5.2.1.4.2.-4.-6.-5.-1$ and $1.5.2.1.4.2.-4.-6.-1.3$ the $\nu_e$ completely decouples which is inconsistent with solar neutrino oscillations. This leaves only the models $1.5.2.1.4.2.-4.-6.-3.-1$ and $1.5.2.1.4.2.-4.-5.-3.-1$. 

We turn to studying the two remaining possibilities. Consider the model $1.5.2.1.4.2.-4.-6.-3.-1$. The Lagrangian interactions read
\be
{\cal L} \supset v \epsilon_2^2 \epsilon_{-4} \left(\nu_{\mu} + \epsilon_{\bar{4}}\nu_{\tau} \right) N_3 + M_* \epsilon_2^2\epsilon_{\bar{4}}^2 N_3 N_3 + \frac{v\mu}{M_*} \left( \epsilon_2 \nu_e N_2 + \nu_{\mu} N_2 + \nu_e N_1 \right) \;.
\ee
We can integrate out the heavy $N_3$ to get
\be
{\cal L} \supset \frac{v^2}{M_*} \epsilon_2^2 \left(\nu_{\mu} + \epsilon_{\bar{4}}\nu_{\tau} \right)^2  + \frac{v\mu}{M_*} \left( \epsilon_2 \nu_e N_2 + \nu_{\mu} N_2 + \nu_e N_1 \right) \;.
\ee
Writing out the mass matrix we can determine the eigenstates. Putting in the values for $\epsilon_{\bar{4}}$ and $\epsilon_2$, setting $\mu=v$ and taking factors of order 1 as exactly 1 we get the eigenvectors
\be
\left( \begin{array}{c} \nu_{e} \\ \nu_{\mu} \\ \nu_{\tau} \\ N_2 \\ N_1 \end{array} \right) =
\left( \begin{array}{c} 0.5 \\ 0.5 \\ 10^{-4} \\ 0.5 \\ 0.5 \end{array} \right) \;,\;
\left( \begin{array}{c} 0.5 \\ 0.5 \\ -10^{-4} \\ -0.5 \\ -0.5 \end{array} \right) \;,\;
\left( \begin{array}{c} 0.5 \\ -0.5 \\ -10^{-4} \\ -0.5 \\ 0.5 \end{array} \right) \;,\;
\left( \begin{array}{c} -0.5 \\ 0.5 \\ -10^{-4} \\ -0.5 \\ 0.5 \end{array} \right) \;,\;
\left( \begin{array}{c} 10^{-10} \\ 10^{-8} \\ -1 \\ 10^{-4} \\ 10^{-5} \end{array} \right) \;,
\ee
with respective eigenvalues, in units of $v^2/M_* \sim 10^{-3} \mathrm{eV}$, $(1,-1,1,-1,10^{-5})$. The scenario is that of 5 Majorana neutrinos: a light one which is mostly $\nu_{\tau}$ and 4 heavy ones that approximately pair up into 2 Dirac ones. However we see that there is very little mixing between $\nu_{\tau}$ and the other neutrinos. This rules these models out in that they can not explain the atmospheric neutrino oscillations. We note that it is not too far from being viable though there may also be constraints from oscillations into sterile neutrinos. 

Consider the model $1.5.2.1.4.2.-4.-5.-3.-1$. Now we find for the eigenvectors 
\be
\left( \begin{array}{c} \nu_{e} \\ \nu_{\mu} \\ \nu_{\tau} \\ N_2 \\ N_1 \end{array} \right) =
\left( \begin{array}{c} -0.03 \\ -0.7 \\ -10^{-4} \\ -0.7 \\ 0 \end{array} \right) \;,\;
\left( \begin{array}{c} -0.03 \\ -0.7 \\ 10^{-4} \\ 0.7 \\ 0 \end{array} \right) \;,\;
\left( \begin{array}{c} -0.2 \\ -0.01 \\ 0.98 \\ -10^{-4} \\ 0 \end{array} \right) \;,\;
\left( \begin{array}{c} 0.97 \\ -0.04 \\ 0.2 \\ 0 \\ 0 \end{array} \right) \;,\;
\left( \begin{array}{c} 0 \\ 0 \\ 0 \\ 0 \\ 1 \end{array} \right) \;,
\ee
with respective eigenvalues $(1,-1,10^{-4},10^{-16},0)$. Again this is a 5 Majorana neutrinos scenario. Now 1 massless Majorana $N_1$ decouples completely. We are left with one heavy approximately Dirac pair which is mostly $\nu_{\mu}$ and two Majorana which are mostly $\nu_e$ and $\nu_{\tau}$ that are light and (essentially) massless respectively. Indeed the mass eigenstate which is mostly $\nu_{e}$ is really too light to produce a big enough mass gap with the massless one. On the positive side there is more substantial mixing between $\nu_{\mu}$ and $\nu_{\tau}$ so that potentially the atmospheric oscillations could be matched. 

These last two scenarios both have their problems, however they are not too far out. For this reason it is worth noting a feature in which they differ from the more phenomenologically attractive model of section \ref{sec:candmod}. The feature is that these models have a singlet $\un_{5}$ in the first and $\un_{6}$ in the second which can play the role of the supersymmetry breaking field in the Giudice-Masiero mechanism ($Y$ in section \ref{sec:muterm}). In the first model we can form the operator in the Kahler potential $\un_{\bar{4}}\un_{\bar{4}}\un_2\un_2\un_5\f_{H_u}\fb_{H_d}$  and in the second model $\un_{\bar{4}}\un_2\un_2\un_6\f_{H_u}\fb_{H_d}$. In this sense these models are more `calculable' than that of section \ref{sec:candmod} since all the needed fields come from the point of $E_8$.

\subsection{$3$-field models}

\subsubsection{Quark sector}

In this section the models are labeled as
\be
\mathrm{\bf Mono\;case}.\fb_{H_d}.\fb_{M1}.\fb_{M2}.\fb_{M3}.\bf{1}_{S1}.\bf{1}_{S2}.\bf{1}_{S3} \;.
\ee

\scriptsize

\bea
\mathrm{Model}\;1.5.2.1.4.-4.-1.2 : 
Y^U = \left( \begin{array}{ccc} \epsilon_{2}\epsilon_{2} & \epsilon_{2}\epsilon_{2}\epsilon_{-4} & \epsilon_{2} \\ \epsilon_{2}\epsilon_{2}\epsilon_{-4} & \epsilon_{2}\epsilon_{2}\epsilon_{-4}\epsilon_{-4} & \epsilon_{2}\epsilon_{-4} \\ \epsilon_{2} & \epsilon_{2}\epsilon_{-4} & 1 \end{array} \right) \;,\;\; 
Y^D = \left( \begin{array}{ccc} \epsilon_{2}\epsilon_{2}\epsilon_{-4} & \epsilon_{2}\epsilon_{-4} & \epsilon_{2} \\ \epsilon_{2}\epsilon_{2}\epsilon_{-4}\epsilon_{-4} & \epsilon_{2}\epsilon_{-4}\epsilon_{-4} & \epsilon_{2}\epsilon_{-4} \\ \epsilon_{2}\epsilon_{-4} & \epsilon_{-4}& 1\end{array} \right) \;.
\eea

\bea
\mathrm{Model}\;1.5.2.1.4.-4.1.2 : 
Y^U = \left( \begin{array}{ccc} \epsilon_{2}\epsilon_{2} & \epsilon_{1}\epsilon_{2}+\epsilon_{2}\epsilon_{2}\epsilon_{-4} & \epsilon_{2} \\ \epsilon_{1}\epsilon_{2}+\epsilon_{2}\epsilon_{2}\epsilon_{-4} & \epsilon_{1}\epsilon_{1}+\epsilon_{1}\epsilon_{2}\epsilon_{-4}+\epsilon_{2}\epsilon_{2}\epsilon_{-4}\epsilon_{-4} & \epsilon_{1}+\epsilon_{2}\epsilon_{-4} \\ \epsilon_{2} & \epsilon_{1}+\epsilon_{2}\epsilon_{-4} & 1\end{array} \right) \;,\;\; \\
Y^D = \left( \begin{array}{ccc} \epsilon_{1}\epsilon_{2}+\epsilon_{2}\epsilon_{2}\epsilon_{-4} & \epsilon_{1}+\epsilon_{2}\epsilon_{-4} & \epsilon_{2} \\ \epsilon_{1}\epsilon_{1}+\epsilon_{1}\epsilon_{2}\epsilon_{-4}+\epsilon_{2}\epsilon_{2}\epsilon_{-4}\epsilon_{-4} & \epsilon_{1}\epsilon_{-4}+\epsilon_{2}\epsilon_{-4}\epsilon_{-4} & \epsilon_{1}+\epsilon_{2}\epsilon_{-4} \\ \epsilon_{1}+\epsilon_{2}\epsilon_{-4} & \epsilon_{-4} & 1\end{array} \right) \;.
\eea

\bea
\mathrm{Model}\;1.5.2.1.4.-2.1.4 : 
Y^U = \left( \begin{array}{ccc} \epsilon_{1}\epsilon_{1}\epsilon_{4}\epsilon_{4} & \epsilon_{1}\epsilon_{1}\epsilon_{4} & \epsilon_{1}\epsilon_{4} \\ \epsilon_{1}\epsilon_{1}\epsilon_{4} & \epsilon_{1}\epsilon_{1} & \epsilon_{1} \\ \epsilon_{1}\epsilon_{4} & \epsilon_{1} & 1\end{array} \right) \;,\;\; 
Y^D = \left( \begin{array}{ccc} \epsilon_{1}\epsilon_{1}\epsilon_{4} & \epsilon_{1} & \epsilon_{1}\epsilon_{4} \\ \epsilon_{1}\epsilon_{1} & \epsilon_{1}\epsilon_{1}\epsilon_{-2} & \epsilon_{1} \\ \epsilon_{1}& \epsilon_{1}\epsilon_{-2} & 1\end{array} \right) \;.
\eea

\bea
\mathrm{Model}\;1.5.2.1.4.1.2.4 : 
Y^U = \left( \begin{array}{ccc} \epsilon_{2}\epsilon_{2}+\epsilon_{1}\epsilon_{2}\epsilon_{4}+\epsilon_{1}\epsilon_{1}\epsilon_{4}\epsilon_{4} & \epsilon_{1}\epsilon_{2}+\epsilon_{1}\epsilon_{1}\epsilon_{4} & \epsilon_{2}+\epsilon_{1}\epsilon_{4} \\ \epsilon_{1}\epsilon_{2}+\epsilon_{1}\epsilon_{1}\epsilon_{4} & \epsilon_{1}\epsilon_{1} & \epsilon_{1} \\ \epsilon_{2}+\epsilon_{1}\epsilon_{4} & \epsilon_{1} & 1\end{array} \right) \;,\;\; \\
Y^D = \left( \begin{array}{ccc} \epsilon_{1}\epsilon_{2}+\epsilon_{1}\epsilon_{1}\epsilon_{4} & \epsilon_{1} & \epsilon_{2}+\epsilon_{1}\epsilon_{4} \\ \epsilon_{1}\epsilon_{1} & 0 & \epsilon_{1} \\ \epsilon_{1} & 0 & 1\end{array} \right) \;.
\eea

\normalsize 

The models $1.5.2.1.4.-4.1.2$ and $1.5.2.1.4.1.2.4$ are ruled out due to a bad neutrino sector (see next section) and bad down quark mixing respectively. The model $1.5.2.1.4.-2.1.4$ is also ruled out due to giving a bad CKM matrix. There remains one model that is $1.5.2.1.4.-4.-1.2$. This model has the same quark sector as the 2-field case and the vev for $\epsilon_{\bar{1}}$ is left as a free parameter.

\subsubsection{Neutrino sector}

In this section the models are labeled as
\be
\mathrm{\bf Mono\;case}.\fb_{H_d}.\fb_{M1}.\fb_{M2}.\fb_{M3}.\bf{1}_{S1}.\bf{1}_{S2}.\bf{1}_{S3}.\bf{1}_{RHN1}.\bf{1}_{RHN2}.\bf{1}_{RHN3} \;.
\ee

\scriptsize

\bea
\mathrm{Model}\;& &1.5.2.1.4.-4.-1.2.-6.-5.-3.\;\;\;\;K = \left( \begin{array}{ccc} \epsilon_{2} & 1 & \epsilon_{-1}\epsilon_{2} \\ \epsilon_{-1}\epsilon_{2} & \epsilon_{-1} & \epsilon_{-1}\epsilon_{-1}\epsilon_{2} \\ 1 & \epsilon_{-1}\epsilon_{-4} & \epsilon_{-1} 
\end{array} \right) \;,\\ 
W &=& \left( \begin{array}{ccc} 0 & 0 & 0 \\ 0 & 0 & 0 \\ 0 & 0 & 0\end{array} \right) \;,\;\;\;M = \left( \begin{array}{ccc} 0 & 0 & 0 \\ 0 & 0 & 0 \\ 0 & 0 & 0\end{array} \right) \;.
\eea

\bea
\mathrm{Model}\;& &1.5.2.1.4.-4.1.2.-6.-5.-3.\;\;\;\;K = \left( \begin{array}{ccc} \epsilon_{2} & 1 & 0 \\ 0 & 0 & 0 \\ 1 & 0 & 0\end{array} \right) \;,\\ 
W &=& \left( \begin{array}{ccc} 0 & 0 & 0 \\ 0 & 0 & 0 \\ 0 & 0 & 0\end{array} \right) \;,\;\;\;M = \left( \begin{array}{ccc} 0 & 0 & 0 \\ 0 & 0 & 0 \\ 0 & 0 & 0\end{array} \right) \;.
\eea

\bea
\mathrm{Model}\;& &1.5.2.1.4.-2.1.4.-6.-5.-3.\;\;\;\;K = \left( \begin{array}{ccc} \epsilon_{1}\epsilon_{4} & 1 & \epsilon_{4} \\ \epsilon_{4} & \epsilon_{-2}\epsilon_{4} & \epsilon_{-2}\epsilon_{4}\epsilon_{4} \\ 1 & \epsilon_{-2} & \epsilon_{-2}\epsilon_{4} \end{array} \right) \;,\\ 
W &=& \left( \begin{array}{ccc} 0 & 0 & 0 \\ 0 & 0 & 0 \\ 0 & 0 & 0\end{array} \right) \;,\;\;\;M = \left( \begin{array}{ccc} 0 & 0 & 0 \\ 0 & 0 & 0 \\ 0 & 0 & 0\end{array} \right) \;.
\eea

\bea
\mathrm{Model}\;& &1.5.2.1.4.1.2.4.-6.-5.-3.\;\;\;\;K = \left( \begin{array}{ccc} \epsilon_{2}+\epsilon_{1}\epsilon_{4} & 1 & \epsilon_{4} \\ \epsilon_{4} & 0 & 0 \\ 1 & 0 & 0\end{array} \right) \;,\\ 
W &=& \left( \begin{array}{ccc} 0 & 0 & 0 \\ 0 & 0 & 0 \\ 0 & 0 & 0\end{array} \right) \;,\;\;\;M = \left( \begin{array}{ccc} 0 & 0 & 0 \\ 0 & 0 & 0 \\ 0 & 0 & 0\end{array} \right) \;.
\eea

\normalsize 

All the models apart from $1.5.2.1.4.-4.-1.2.-6.-5.-3$ are ruled out because their respective quark sectors are incompatible with observation. The model $1.5.2.1.4.-4.1.2.-6.-5.-3$ also has one neutrino decoupled which is ruled out from neutrino mixing constraints.

\section{Models with vector pairs of singlets}
\label{sec:vector}

In this appendix we study models where we allow some of the singlets to come in vector pairs. Of course this is slightly unattractive in terms of model building since we expect a vector-like pair to gain a UV scale mass and to not appear in a low energy effective theory. For this reason these models are relegated to the appendix. 

We find that the only phenomenologically relevant cases are constructed from the 3-field cases studied in the previous section but with extensions by some conjugates of the singlets already present. This means that the matter curves are the same single combinations $1.5.2.1.4$ as in the previous sections and the singlets are always $\bf{1}_{1}$, $\bf{1}_{2}$, $\bf{1}_{4}$ and their conjugates. In particular this implies that the global $U(1)$ symmetry associated to $t_5$ that was present in the model of section \ref{sec:candmod} is also present for these models. This symmetry forbids a $\mu$ term and proton decay operators.

There are certainly new models within this class that can satisfy the required phenomenological constraints. We discuss some examples in section \ref{sec:vecgoodmodel}. Although they offer alternatives to the model of section \ref{sec:candmod} they do not carry many advantages over it. There are two differences worth mentioning thought. The first is the fact that it is possible to have quark Yukawas that have 3 independent parameters in their entries (in the chiral case there were only ever 2 independent parameters). This means that it is possible to tune the Yukawas to match the quark masses and mixing more precisely. This is not much of an advantage given that the quark masses and mixing can already by matched quite well by the model in section \ref{sec:candmod}. The second difference is that, like in the 2-field models of section \ref{sec:2fsemiviab}, there are candidate singlets for the Giudice-Masiero field.

\subsection{3-field models}
\label{sec:3fvector}

\subsubsection{The quark sector}
\label{sec:3fquarkvector}

In this section the models are labeled as
\be
\mathrm{\bf Mono\;case}.\fb_{H_d}.\fb_{M1}.\fb_{M2}.\fb_{M3}.\bf{1}_{S1}.\bf{1}_{S2}.\bf{1}_{S3} \;.
\ee

\scriptsize

\be
\mathrm{Model}\;1.5.2.1.4.-4.-2.2 : 
Y^U = \left( \begin{array}{ccc} \epsilon_{2}\epsilon_{2} & \epsilon_{2}\epsilon_{2}\epsilon_{-4} & \epsilon_{2} \\ \epsilon_{2}\epsilon_{2}\epsilon_{-4} & \epsilon_{2}\epsilon_{2}\epsilon_{-4}\epsilon_{-4} & \epsilon_{2}\epsilon_{-4} \\ \epsilon_{2}  & \epsilon_{2}\epsilon_{-4}  & 1 \end{array} \right) \;,\;\;
Y^D = \left( \begin{array}{ccc} \epsilon_{2}\epsilon_{2}\epsilon_{-4} & \epsilon_{2}\epsilon_{-4} & \epsilon_{2}  \\ \epsilon_{2}\epsilon_{2}\epsilon_{-4}\epsilon_{-4} & \epsilon_{2}\epsilon_{-4}\epsilon_{-4} & \epsilon_{2}\epsilon_{-4} \\ \epsilon_{2}\epsilon_{-4}  & \epsilon_{-4} & 1 \end{array} \right) \;. \label{qrk1}
\ee

\be
\mathrm{Model}\;1.5.2.1.4.-4.1.4 : 
Y^U = \left( \begin{array}{ccc} \epsilon_{1}\epsilon_{1}\epsilon_{4}\epsilon_{4} & \epsilon_{1}\epsilon_{1}\epsilon_{4} & \epsilon_{1}\epsilon_{4} \\ \epsilon_{1}\epsilon_{1}\epsilon_{4} & \epsilon_{1}\epsilon_{1} & \epsilon_{1} \\ \epsilon_{1}\epsilon_{4} & \epsilon_{1}  & 1 \end{array} \right) \;,\;\;
Y^D = \left( \begin{array}{ccc} \epsilon_{1}\epsilon_{1}\epsilon_{4}  & \epsilon_{1} & \epsilon_{1}\epsilon_{4}\\ \epsilon_{1}\epsilon_{1} & \epsilon_{1}\epsilon_{-4}& \epsilon_{1} \\ \epsilon_{1} & \epsilon_{-4}& 1 \end{array} \right) \;.
\ee

\be
\mathrm{Model}\;1.5.2.1.4.-4.2.4 : 
Y^U = \left( \begin{array}{ccc} \epsilon_{2}\epsilon_{2} & \epsilon_{2}\epsilon_{2}\epsilon_{-4} & \epsilon_{2} \\ \epsilon_{2}\epsilon_{2}\epsilon_{-4} & \epsilon_{2}\epsilon_{2}\epsilon_{-4}\epsilon_{-4} & \epsilon_{2}\epsilon_{-4} \\ \epsilon_{2}  & \epsilon_{2}\epsilon_{-4}  & 1 \end{array} \right) \;,\;\;
Y^D = \left( \begin{array}{ccc} \epsilon_{2}\epsilon_{2}\epsilon_{-4} & \epsilon_{2}\epsilon_{-4} & \epsilon_{2}  \\ \epsilon_{2}\epsilon_{2}\epsilon_{-4}\epsilon_{-4} & \epsilon_{2}\epsilon_{-4}\epsilon_{-4} & \epsilon_{2}\epsilon_{-4} \\ \epsilon_{2}\epsilon_{-4}  & \epsilon_{-4} & 1 \end{array} \right) \;. \label{qrk2}
\ee

\be
\mathrm{Model}\;1.5.2.1.4.-2.1.2 : 
Y^U = \left( \begin{array}{ccc} \epsilon_{2}\epsilon_{2} & \epsilon_{1}\epsilon_{2} & \epsilon_{2} \\ \epsilon_{1}\epsilon_{2} & \epsilon_{1}\epsilon_{1}& \epsilon_{1} \\ \epsilon_{2} & \epsilon_{1} & 1 \end{array} \right) \;,\;\; 
Y^D = \left( \begin{array}{ccc} \epsilon_{1}\epsilon_{2} & \epsilon_{1} & \epsilon_{2}\\ \epsilon_{1}\epsilon_{1} & \epsilon_{1}\epsilon_{1}\epsilon_{-2} & \epsilon_{1} \\ \epsilon_{1} & \epsilon_{1}\epsilon_{-2} & 1 \end{array} \right) \;. \label{qrk3}
\ee

\be
\mathrm{Model}\;1.5.2.1.4.-1.1.2 : 
Y^U = \left( \begin{array}{ccc} \epsilon_{2}\epsilon_{2} & \epsilon_{1}\epsilon_{2} & \epsilon_{2} \\ \epsilon_{1}\epsilon_{2} & \epsilon_{1}\epsilon_{1} & \epsilon_{1} \\ \epsilon_{2} & \epsilon_{1} & 1 \end{array} \right) \;,\;\; 
Y^D = \left( \begin{array}{ccc} \epsilon_{1}\epsilon_{2} & \epsilon_{1} & \epsilon_{2} \\ \epsilon_{1}\epsilon_{1} & 0 & \epsilon_{1} \\ \epsilon_{1} & 0 & 1 \end{array} \right) \;.
\ee

\be
\mathrm{Model}\;1.5.2.1.4.-1.1.4 : 
Y^U = \left( \begin{array}{ccc} \epsilon_{1}\epsilon_{1}\epsilon_{4}\epsilon_{4} & \epsilon_{1}\epsilon_{1}\epsilon_{4} & \epsilon_{1}\epsilon_{4} \\ \epsilon_{1}\epsilon_{1}\epsilon_{4} & \epsilon_{1}\epsilon_{1} & \epsilon_{1} \\ \epsilon_{1}\epsilon_{4} & \epsilon_{1} & 1 \end{array} \right) \;,\;\; 
Y^D = \left( \begin{array}{ccc} \epsilon_{1}\epsilon_{1}\epsilon_{4} & \epsilon_{1} & \epsilon_{1}\epsilon_{4} \\ \epsilon_{1}\epsilon_{1} & 0 & \epsilon_{1} \\ \epsilon_{1} & 0 & 1 \end{array} \right) \;.
\ee

\normalsize 

The models $1.5.2.1.4.-4.-2.2$ and $1.5.2.1.4.-4.2.4$ simply recreate the quark sector of the analogous (chiral) 2-field model. Of the four new models, only $1.5.2.1.4.-2.1.2$ can recreate a realistic quark sector (see section \ref{sec:vecgoodmodel}). The rest give rise to too large mixing.

\subsubsection{The neutrino sector}
\label{sec:3fneutvector}

In this section the models are labeled as
\be
\mathrm{\bf Mono\;case}.\fb_{H_d}.\fb_{M1}.\fb_{M2}.\fb_{M3}.\bf{1}_{S1}.\bf{1}_{S2}.\bf{1}_{S3}.\bf{1}_{\bf{RHN1}}.\bf{1}_{\bf{RHN2}}.\bf{1}_{\bf{RHN3}} \;.
\ee

We display all the relevant neutrino models that are based on the phenomenologically viable quark models presented in the previous section, i.e. (\ref{qrk1}), (\ref{qrk2}) and (\ref{qrk3}). 

\scriptsize

\bea
\mathrm{Model}\;& &1.5.2.1.4.-4.-2.2.-6.-5.-1.\;\;\;\;K = \left( \begin{array}{ccc} 0 & 0 & 0 \\ 0 & 0 & 0 \\ 1 & \epsilon_{-2} & 0 \end{array} \right) \;,\\ 
W &=& \left( \begin{array}{ccc} 0 & \epsilon_{2}\epsilon_{2}\epsilon_{-4}\epsilon_{-4} & \epsilon_{2}\epsilon_{2}\epsilon_{-4} \\ 0 & 0 & 0 \\ 0 & 0 & 0\end{array} \right) \;,\;\;\;M = \left( \begin{array}{ccc} \epsilon_{2}\epsilon_{2}\epsilon_{-4}\epsilon_{-4} & 0 & 0 \\ 0 & 0 & 0 \\ 0 & 0 & 0\end{array} \right) \;.
\eea

\bea
\mathrm{Model}\;& &1.5.2.1.4.-4.-2.2.-6.-3.-1.\;\;\;\;K = \left( \begin{array}{ccc} 0 & 0 & 0 \\ \epsilon_{2} & 1 & 0 \\ 1 & \epsilon_{-2} & 0 \end{array} \right) \;,\\ 
W &=& \left( \begin{array}{ccc} 0 & \epsilon_{2}\epsilon_{2}\epsilon_{-4}\epsilon_{-4} & \epsilon_{2}\epsilon_{2}\epsilon_{-4} \\ 0 & 0 & 0 \\ 0 & 0 & 0\end{array} \right) \;,\;\;\;M = \left( \begin{array}{ccc} \epsilon_{2}\epsilon_{2}\epsilon_{-4}\epsilon_{-4} & 0 & 0 \\ 0 & 0 & 0 \\ 0 & 0 & 0\end{array} \right) \;.
\eea

\bea
\mathrm{Model}\;& &1.5.2.1.4.-4.-2.2.-5.-3.-1.\;\;\;\;K = \left( \begin{array}{ccc} 0 & 0 & 0 \\ \epsilon_{2} & 1 & 0 \\ 0 & 0 & 0\end{array} \right) \;,\\ 
W &=& \left( \begin{array}{ccc} 0 & \epsilon_{2}\epsilon_{2}\epsilon_{-4}\epsilon_{-4} & \epsilon_{2}\epsilon_{2}\epsilon_{-4} \\ 0 & 0 & 0 \\ 0 & 0 & 0\end{array} \right) \;,\;\;\;M = \left( \begin{array}{ccc} \epsilon_{2}\epsilon_{2}\epsilon_{-4}\epsilon_{-4} & 0 & 0 \\ 0 & 0 & 0 \\ 0 & 0 & 0\end{array} \right) \;.
\eea

\bea
\mathrm{Model}\;& &1.5.2.1.4.-4.2.4.-6.-3.-1.\;\;\;\;K = \left( \begin{array}{ccc} 0 & 0 & 0 \\ \epsilon_{2} & 1 & \epsilon_{4} \\ 1 & 0 & 0\end{array} \right) \;,\\ 
W &=& \left( \begin{array}{ccc} 0 & \epsilon_{2}\epsilon_{2}\epsilon_{-4}\epsilon_{-4} & \epsilon_{2}\epsilon_{2}\epsilon_{-4} \\ 0 & 0 & 0 \\ 0 & 0 & 0\end{array} \right) \;,\;\;\;M = \left( \begin{array}{ccc} \epsilon_{2}\epsilon_{2}\epsilon_{-4}\epsilon_{-4} & 0 & 0 \\ 0 & 0 & 0 \\ 0 & 0 & 0\end{array} \right) \;.
\eea

\bea
\mathrm{Model}\;& &1.5.2.1.4.-4.2.4.-5.-3.-1.\;\;\;\;K = \left( \begin{array}{ccc} 0 & 0 & 0 \\ \epsilon_{2} & 1 & \epsilon_{4}  \\ \epsilon_{4} & 0 & 0\end{array} \right) \;,\\ 
W &=& \left( \begin{array}{ccc} 0 & \epsilon_{2}\epsilon_{2}\epsilon_{-4}\epsilon_{-4} & \epsilon_{2}\epsilon_{2}\epsilon_{-4} \\ 0 & 0 & 0 \\ 0 & 0 & 0\end{array} \right) \;,\;\;\;M = \left( \begin{array}{ccc} \epsilon_{2}\epsilon_{2}\epsilon_{-4}\epsilon_{-4} & 0 & 0 \\ 0 & 0 & 0 \\ 0 & 0 & 0\end{array} \right) \;.
\eea

\bea
\mathrm{Model}\;& &1.5.2.1.4.-2.1.2.-6.-5.4.\;\;\;\;K = \left( \begin{array}{ccc} 0 & 0 & 0 \\ 0 & 0 & 0 \\ 1 & \epsilon_{-2} & 0\end{array} \right) \;,\\ 
W &=& \left( \begin{array}{ccc} 0 & \epsilon_{1}\epsilon_{1}\epsilon_{-2} & \epsilon_{1} \\ 0 & 0 & 0 \\ 0 & 0 & 0\end{array} \right) \;,\;\;\;M = \left( \begin{array}{ccc} \epsilon_{1}\epsilon_{1}\epsilon_{-2}\epsilon_{-2} & 0 & 0 \\ 0 & 0 & 0 \\ 0 & 0 & 0\end{array} \right) \;.
\eea

\bea
\mathrm{Model}\;& &1.5.2.1.4.-2.1.2.-6.-3.4.\;\;\;\;K = \left( \begin{array}{ccc} 0 & 0 & 0 \\ \epsilon_{2} & 1 & 0 \\ 1 & \epsilon_{-2} & 0\end{array} \right) \;,\\ 
W &=& \left( \begin{array}{ccc} 0 & \epsilon_{1}\epsilon_{1}\epsilon_{-2} & \epsilon_{1} \\ 0 & 0 & 0 \\ 0 & 0 & 0\end{array} \right) \;,\;\;\;M = \left( \begin{array}{ccc} \epsilon_{1}\epsilon_{1}\epsilon_{-2}\epsilon_{-2} & 0 & 0 \\ 0 & 0 & 0 \\ 0 & 0 & 0\end{array} \right) \;.
\eea

\bea
\mathrm{Model}\;& &1.5.2.1.4.-2.1.2.-5.-3.4.\;\;\;\;K = \left( \begin{array}{ccc} 0 & 0 & 0 \\ \epsilon_{2} & 1 & 0 \\ 0 & 0 & 0\end{array} \right) \;,\\ 
W &=& \left( \begin{array}{ccc} 0 & \epsilon_{1}\epsilon_{1}\epsilon_{-2} & \epsilon_{1} \\ 0 & 0 & 0 \\ 0 & 0 & 0\end{array} \right) \;,\;\;\;M = \left( \begin{array}{ccc} \epsilon_{1}\epsilon_{1}\epsilon_{-2}\epsilon_{-2} & 0 & 0 \\ 0 & 0 & 0 \\ 0 & 0 & 0\end{array} \right) \;.
\eea

\normalsize

All of these models can give rise to semi-viable models. We study the more attractive possibilities in section \ref{sec:vecgoodmodel}.

\subsection{4-field models}
\label{sec:4fvector}

\subsubsection{The quark sector}
\label{sec:4fquarkvector}

In this section the models are labeled as
\be
\mathrm{\bf Mono\;case}.\fb_{H_d}.\fb_{M1}.\fb_{M2}.\fb_{M3}.\bf{1}_{S1}.\bf{1}_{S2}.\bf{1}_{S3}.\bf{1}_{S4} \;.
\ee

The realistic 4-field quark sector models are
\scriptsize
\bea
& &1.5.2.1.4.-4.-2.-1.2 \;,  \nn \\
& &1.5.2.1.4.-4.-2.1.2  \;,  \nn \\
& &1.5.2.1.4.-4.-2.2.4  \;,  \nn \\
& &1.5.2.1.4.-4.-1.1.2  \;,  \nn \\
& &1.5.2.1.4.-4.-1.1.4  \;,  \nn \\
& &1.5.2.1.4.-4.-1.2.4  \;,   \nn \\
& &1.5.2.1.4.-4.1.2.4   \;, \nn \\
& &1.5.2.1.4.-2.-1.1.2  \;,  \nn \\
& &1.5.2.1.4.-2.-1.1.4  \;,  \nn \\
& &1.5.2.1.4.-2.1.2.4   \;,  \nn \\
& &1.5.2.1.4.-1.1.2.4   \;. 
\eea
\normalsize
The 4-field cases that include a 3-field case which had full Yukawa matrices just reproduce the same Yukawas as the corresponding 3-field case. The exception to this is that since $\bf{1}_{1}=\bf{1}_{2}\bar{\bf{1}}_{4}$ in the cases where all of these singlets (or their conjugates) are involved there are two relevant contributions to the Yukawas. Below we display the only cases where the Yukawas differ from the corresponding 3-field case other than through the extra contribution from the new gauge invariant combination just discussed, i.e. we show the cases where Yukawa entries that vanished in the 3-field case are now non-vanishing.
\scriptsize

\be
\mathrm{Model}\;1.5.2.1.4.-4.-1.1.2 : 
Y^U = \left( \begin{array}{ccc} \epsilon_{2}\epsilon_{2} & \epsilon_{1}\epsilon_{2} & \epsilon_{2} \\ \epsilon_{1}\epsilon_{2} & \epsilon_{1}\epsilon_{1} & \epsilon_{1} \\ \epsilon_{2} & \epsilon_{1} & 1 \end{array} \right) \;,\;\; 
Y^D = \left( \begin{array}{ccc} \epsilon_{1}\epsilon_{2} & \epsilon_{1} & \epsilon_{2} \\ \epsilon_{1}\epsilon_{1} & \epsilon_{1}\epsilon_{-4} & \epsilon_{1} \\ \epsilon_{1} & \epsilon_{-4} & 1 \end{array} \right) \;.
\ee

\be
\mathrm{Model}\;1.5.2.1.4.-4.-1.1.4 : 
Y^U = \left( \begin{array}{ccc} \epsilon_{1}\epsilon_{1}\epsilon_{4}\epsilon_{4} & \epsilon_{1}\epsilon_{1}\epsilon_{4} & \epsilon_{1}\epsilon_{4} \\ \epsilon_{1}\epsilon_{1}\epsilon_{4} & \epsilon_{1}\epsilon_{1} & \epsilon_{1}  \\ \epsilon_{1}\epsilon_{4} & \epsilon_{1} & 1\end{array} \right) \;,\;\; 
Y^D = \left( \begin{array}{ccc} \epsilon_{1}\epsilon_{1}\epsilon_{4} & \epsilon_{1} & \epsilon_{1}\epsilon_{4} \\ \epsilon_{1}\epsilon_{1}& \epsilon_{1}\epsilon_{-4} & \epsilon_{1}\\ \epsilon_{1} & \epsilon_{-4} & 1 \end{array} \right) \;.
\ee

\be
\mathrm{Model}\;1.5.2.1.4.-2.-1.1.2 : 
Y^U = \left( \begin{array}{ccc} \epsilon_{2}\epsilon_{2}& \epsilon_{1}\epsilon_{2}& \epsilon_{2} \\ \epsilon_{1}\epsilon_{2} & \epsilon_{1}\epsilon_{1} & \epsilon_{1} \\ \epsilon_{2} & \epsilon_{1} & 1 \end{array} \right) \;,\;\;
Y^D = \left( \begin{array}{ccc} \epsilon_{1}\epsilon_{2} & \epsilon_{1} & \epsilon_{2} \\ \epsilon_{1}\epsilon_{1} & \epsilon_{1}\epsilon_{1}\epsilon_{-2} & \epsilon_{1} \\ \epsilon_{1} & \epsilon_{1}\epsilon_{-2} & 1 \end{array} \right) \;.
\ee

\be
\mathrm{Model}\;1.5.2.1.4.-2.-1.1.4 : 
Y^U = \left( \begin{array}{ccc} \epsilon_{1}\epsilon_{1}\epsilon_{4}\epsilon_{4} & \epsilon_{1}\epsilon_{1}\epsilon_{4} & \epsilon_{1}\epsilon_{4} \\ \epsilon_{1}\epsilon_{1}\epsilon_{4} & \epsilon_{1}\epsilon_{1} & \epsilon_{1} \\ \epsilon_{1}\epsilon_{4} & \epsilon_{1} & 1 \end{array} \right) \;,\;\;
Y^D = \left( \begin{array}{ccc} \epsilon_{1}\epsilon_{1}\epsilon_{4} & \epsilon_{1} & \epsilon_{1}\epsilon_{4} \\ \epsilon_{1}\epsilon_{1} & \epsilon_{1}\epsilon_{1}\epsilon_{-2} & \epsilon_{1} \\ \epsilon_{1} & \epsilon_{1}\epsilon_{-2} & 1 \end{array} \right) \;.
\ee

\normalsize 

We do not go on to consider 5-field and 6-field cases since they do not add any important new features. Also we do not go on to consider the 4-field neutrino sector since there are too many viable models to display. They can all be built as extensions of the 3-field models in the previous section.

\subsection{Example candidate models}
\label{sec:vecgoodmodel}

Allowing for vector-like pairs in the singlet sector opens up a number of new phenomenologically relevant models. In this section we study two examples.\footnote{Both of the examples have local singlet candidates for a Giudice-Masiero field and in that sense differ from the model presented in section \ref{sec:candmod}.}

\subsubsection{Example 1}
\label{sec:vecexa1}

The first example is the model $1.5.2.1.4.-4.2.4.-6.-3.-1$. This has the same quark sector as the model in section \ref{sec:candmod} and so the same analysis applies. The neutrino sector however is different and reads
\bea
K = \left( \begin{array}{ccc} 0 & 0 & 0 \\ \epsilon_{2} & 1 & \epsilon_{4} \\ 1 & 0 & 0\end{array} \right) \;,\; 
W = \left( \begin{array}{ccc} 0 & \epsilon_{2}^2\epsilon_{-4}^2 & \epsilon_{2}^2\epsilon_{-4} \\ 0 & 0 & 0 \\ 0 & 0 & 0\end{array} \right) \;,\;M = \left( \begin{array}{ccc} \epsilon_{2}^2\epsilon_{-4}^2 & 0 & 0 \\ 0 & 0 & 0 \\ 0 & 0 & 0\end{array} \right) \;.
\eea
After integrating out $N_3$ we find the effective Lagrangian
\be
{\cal L} \supset \frac{v^2}{M_*} \left( \epsilon_2^2 \epsilon_{-4}^2 \nu_{\mu}^2 + 2\epsilon_{2}^2\epsilon_{-4}\nu_{\mu}\nu_{\tau} + \epsilon_{2}^2\nu_{\tau}^2 \right)  + \frac{v\mu}{M_*} \left[ \left(\epsilon_2 \nu_e + \nu_{\mu} + \epsilon_{-4} \nu_{\tau} \right) N_2 + \nu_e N_1 \right] \;.
\ee
Taking $\mu=v$ the resulting the mass eigenvectors are
\be
\left( \begin{array}{c} \nu_{e} \\ \nu_{\mu} \\ \nu_{\tau} \\ N_2 \\ N_1 \end{array} \right) =
\left( \begin{array}{c} -0.4 \\ -0.6 \\ -0.1 \\ -0.6 \\ -0.4 \end{array} \right) \;,\;
\left( \begin{array}{c} -0.4 \\ -0.6 \\ -0.1 \\ 0.6 \\ 0.4 \end{array} \right) \;,\;
\left( \begin{array}{c} 0.6 \\ -0.4 \\ -0.1 \\ -0.4 \\ 0.6 \end{array} \right) \;,\;
\left( \begin{array}{c} -0.6 \\ 0.4 \\ 0.1 \\ -0.4 \\ 0.6 \end{array} \right) \;,\;
\left( \begin{array}{c} 10^{-8} \\ 0.2 \\ -1 \\ 0.001 \\ 10^{-5} \end{array} \right) \;,
\ee
with respective eigenvalues, in units of $v^2/M_* \sim 10^{-3} \mathrm{eV}$, $(1,-1,1,-1,10^{-3})$. The model is similar to that studied in section \ref{sec:2fns}. It has two approximately Dirac neutrinos and one Majorana one. However now the light Majorana neutrino is not too light but more importantly there is order one mixing between the neutrinos. In many ways this model offers an attractive alternative to that presented in section \ref{sec:candmod}.

\subsubsection{Example 2}
\label{sec:vecexa2}

The second example has a new quark sector compared to the model studied in section \ref{sec:candmod}. Although the quark sector is not as phenomenologically attractive we still present it as an interesting alternative. The model we consider is denoted as $1.5.2.1.4.-2.1.2.-6.-3.4$ in section \ref{sec:3fneutvector}. After interchanging two of the $\te$ and two of the $\f$ curves we have the Yukawa matrices
\be
Y^U = \left( \begin{array}{ccc} \epsilon_{1}^2 & \epsilon_{1}\epsilon_{2} & \epsilon_{1} \\ \epsilon_{1}\epsilon_{2} & \epsilon_{2}^2 & \epsilon_{2} \\ \epsilon_{1} & \epsilon_{2} & 1 \end{array} \right) \;,\;\; 
Y^D = \left( \begin{array}{ccc} \epsilon_{1}^2\epsilon_{-2} & \epsilon_{1}^2 & \epsilon_{1}\\ \epsilon_{1}  & \epsilon_{1}\epsilon_{2} & \epsilon_{2} \\ \epsilon_{1}\epsilon_{-2} & \epsilon_{1} & 1 \end{array} \right) \;.
\ee
Setting $\epsilon_{1}=\lambda^3$ and $\epsilon_{2}=\lambda^2$ and repeating the analysis of section \ref{sec:goodmodquark} we recover exactly the Wolfenstein CKM matrix. The quark masses however are in the ratios
\bea
& &\mathrm{Up\;quarks}\;\;\lambda^6:\lambda^4:1 \;, \\
& &\mathrm{Down\;quarks}\;\;\lambda^6\epsilon_{-2}:\lambda^5:1 \;.
\eea
The down-type ratios are not very attractive in that the strange and down quarks are too light. This certainly also imposes that $\epsilon_{-2}$ should not be too small.

The neutrino sector reads (after interchanging the two $\f$ curves)
\bea
D_K = \left( \begin{array}{ccc} 0 & 0 & 0 \\ 1 & \epsilon_{2} & 0 \\ \epsilon_{-2} & 1 & 0\end{array} \right) \;,\;D_W = \left( \begin{array}{ccc}  \epsilon_{1}^2\epsilon_{-2} & 0 & \epsilon_{1} \\ 0 & 0 & 0 \\ 0 & 0 & 0\end{array} \right) \;,\;M = \left( \begin{array}{ccc} \epsilon_{1}^2\epsilon_{-2}^2 & 0 & 0 \\ 0 & 0 & 0 \\ 0 & 0 & 0\end{array} \right) \;.
\eea
After integrating out $N_3$ we find an effective Lagrangian
\be
{\cal L} \supset \frac{v^2}{M_*} \left( \epsilon_1^2 \nu_{e}^2 + 2\frac{\epsilon_{1}}{\epsilon_{-2}}\nu_e\nu_{\tau} + \frac{1}{\epsilon_{-2}^2}\nu_{\tau}^2 \right)  + \frac{v\mu}{M_*} \left[ \left(\nu_e + \epsilon_2 \nu_{\mu} \right) N_2 + \left(\nu_{\mu} + \epsilon_{-2}\nu_e \right) N_1 \right] \;.
\ee
Taking $\epsilon_{-2}=\lambda$ and $\mu=v$, the mass matrix has the following eigenvectors 
\be
\left( \begin{array}{c} \nu_{e} \\ \nu_{\mu} \\ \nu_{\tau} \\ N_2 \\ N_1 \end{array} \right) =
\left( \begin{array}{c} 0.5 \\ 0.5 \\ 0.02 \\ 0.5 \\ 0.5 \end{array} \right) \;,\;
\left( \begin{array}{c} -0.5 \\ -0.5 \\ 0.02 \\ 0.5 \\ 0.5 \end{array} \right) \;,\;
\left( \begin{array}{c} -0.5 \\ 0.5 \\ -0.02 \\ -0.5 \\ 0.5 \end{array} \right) \;,\;
\left( \begin{array}{c} 0.5 \\ -0.5 \\ -0.02 \\ -0.5 \\ 0.5 \end{array} \right) \;,\;
\left( \begin{array}{c} 10^{-17} \\ 10^{-18} \\ 1 \\ -0.04 \\ 0.002 \end{array} \right) \;,
\ee
with respective eigenvalues, in units of $v^2/M_* \sim 10^{-3} \mathrm{eV}$, $(1.1,-1.1,0.9,-0.9,10^{-16})$. The model is very similar to that studied in section \ref{sec:2fns}. The advantage is that there is more substantial mixing with $\nu_{\tau}$.

Overall this scenario is not as phenomenologically attractive as that presented in section \ref{sec:candmod} but forms an interesting alternative.

\section{The singlet vevs and supersymmetry}
\label{sec:susy}

The vevs of the singlets are taken as input parameters since determining them dynamically is a global issue that does not decouple from moduli stabilisation. However there is some tension between the singlet vevs and supersymmetry that we briefly discuss in this appendix. We discuss the model given in section \ref{sec:candmod}. The model has 3 singlets $X_1$, $X_2$ and $X_3$ and their charges are given in table \ref{tab:goodmodel}. 

The singlets appear in the D-terms associated to the $U(1)$s. We can take the $U(1)$s to be associated to $\{t_1,t_2\}$, $t_3$ and $t_4$ so that the D-terms read 
\bea
D_1 &=& |X_2|^2 - |X_3|^2 + \xi_1 \;, \\
D_2 &=& -|X_1|^2 + |X_3|^2 + \xi_2 \;, \\
D_3 &=& |X_1|^2 - |X_2|^2 + \xi_3 \;. \label{dterms}
\eea
The FI terms $\xi_i$ are moduli dependent. The moduli should adjust themselves so that $\xi_1+\xi_2+\xi_3=0$ which is a necessary condition for supersymmetry. As well as the D-terms there are the F-terms that arise from the superpotential
\be
W \supset \lambda X_1 X_2 X_3 \;. \label{singletsuper}
\ee
Straightforwardly from (\ref{singletsuper}) it follows that if at least 2 of the singlets have a vev, which is the case in our model, some F-terms are non-vanishing. This system has a supersymmetric minimum with 2 vanishing singlet vevs and 1 vanishing FI term. Since we require 3 non-vanishing singlet vevs such a minimum is not viable.

Of course as we have stated the vacuum depends on the global completion of the model. In particular there may be more terms in the superpotential, such as instantonic ones, that involve the singlets $X_i$. It is possible however to alter the model so that a supersymmetric minimum with non-vanishing vevs can be found with only local ingredients. If we do not turn on flux along the singlet curves their conjugate partners are also present and the corresponding mass terms too. Denoting the conjugate singlets $\tilde{X}_i$, the D-terms and superpotential now read
\bea
D_1 &=& |X_2|^2 - |\tilde{X}_2|^2 - |X_3|^2 + |\tilde{X}_3|^2 + \xi_1 \;, \\
D_2 &=& -|X_1|^2 + |\tilde{X}_1|^2 + |X_3|^2 - |\tilde{X}_3|^2 + \xi_2 \;, \\
D_3 &=& |X_1|^2 - |\tilde{X}_1|^2 - |X_2|^2 + |\tilde{X}_2|^2+ \xi_3 \;, \\
W &\supset& \lambda X_1 X_2 X_3 + M_1 X_1 \tilde{X}_1 + M_2 X_2 \tilde{X}_2 + M_3 X_3 \tilde{X}_3 +\lambda' \tilde{X}_1 \tilde{X}_2 \tilde{X}_3 \;. \label{pairsuper}
\eea
It is simple to show that this system has a supersymmetric minimum with non-vanishing singlet vevs. Setting $M_1=M_2=M_3=M$, $\lambda=\lambda'=1$, and $\xi_1=\xi_3=M^2=-\half \xi_2$, we have the solution 
\bea
|X_1|^2 &\simeq& \frac35 M^2 \;, \;\; |X_2|^2 \simeq M^2 \;, \;\; |X_3|^2 \simeq \frac85 M^2 \;, \nn \\
|\tilde{X}_1|^2 &\simeq& \frac53 M^2 \;, \;\; |\tilde{X}_2|^2 \simeq M^2 \;, \;\; |\tilde{X}_3|^2 \simeq \frac58 M^2 \;.
\eea
By taking $M$ small we can consistently keep all the fields in the effective theory below the cutoff scale. 

Although this is a supersymmetric solution the vevs for the $\tilde{X}_i$ ruin the Yukawa coupling calculations. Indeed we have that the vev $X_1\simeq X_2$ which is not consistent with the vevs taken in section \ref{sec:candmod}. Further, since $\tilde{X}_3$ has the same quantum numbers as $X_1 X_2$, its vev also gives too large Yukawa entries. It is possible to pick the masses and FI term appropriately to remedy this situation. If we take
\bea
\lambda&=&\lambda' \;, \;\; M_1=4 \times 10^{-5/2}\lambda M_*\;,\;\; M_2=4 \times 10^{-3/2} \lambda M_*\;,\;\; M_3=\lambda M_*\;, \nn \\
\xi_1 &=& 0.1 M_*^2 \;,\;\; \xi_2 = -0.4 M_*^2 \;,\;\; \xi_3 = -(\xi_1+\xi_2)\;.
\eea
We find the solutions
\bea
|X_1| &\simeq& 0.2 M_* \;, \;\; |X_2| \simeq 0.047 M_*\;, \;\; |X_3| \simeq 0.17 M_*\;, \nn \\
|\tilde{X}_1| &\simeq& 0.64 M_*\;, \;\; |\tilde{X}_2| \simeq 0.27 M_*\;, \;\; |\tilde{X}_3| \simeq 0.009 M_*\;.
\eea
These match the vevs taken for phenomenological motivations in section \ref{sec:candmod}. Note that $X_1 X_2 \simeq \tilde{X}_3 M_*$ and so the vev of $\tilde{X}_3$ leaves the Yukawa matrices calculations unaffected. Again by taking $\lambda$ small we can remain within the validity of the effective field theory although 2 of the  $\tilde{X}_i$ have relatively large vevs.



\begin{thebibliography}{99}

\bibitem{Donagi:2008ca}
  R.~Donagi and M.~Wijnholt,
  ``Model Building with F-Theory,''
  arXiv:0802.2969 [hep-th].

\bibitem{Beasley:2008dc}
  C.~Beasley, J.~J.~Heckman and C.~Vafa,
  ``GUTs and Exceptional Branes in F-theory - I,''
  JHEP {\bf 0901} (2009) 058
  [arXiv:0802.3391 [hep-th]].

\bibitem{Beasley:2008kw}
  C.~Beasley, J.~J.~Heckman and C.~Vafa,
  ``GUTs and Exceptional Branes in F-theory - II: Experimental Predictions,''
  JHEP {\bf 0901} (2009) 059
  [arXiv:0806.0102 [hep-th]].

\bibitem{Donagi:2008kj}
  R.~Donagi and M.~Wijnholt,
  ``Breaking GUT Groups in F-Theory,''
  arXiv:0808.2223 [hep-th].

\bibitem{Aldazabal:2000sa}
  G.~Aldazabal, L.~E.~Ibanez, F.~Quevedo and A.~M.~Uranga,
  ``D-branes at singularities: A bottom-up approach to the string  embedding of
  the standard model,''
  JHEP {\bf 0008} (2000) 002
  [arXiv:hep-th/0005067].

\bibitem{Hayashi:2008ba}
  H.~Hayashi, R.~Tatar, Y.~Toda, T.~Watari and M.~Yamazaki,
  ``New Aspects of Heterotic--F Theory Duality,''
  Nucl.\ Phys.\  B {\bf 806} (2009) 224
  [arXiv:0805.1057 [hep-th]].
  
  \bibitem{08052943}
  L.~Aparicio, D.~G.~Cerdeno and L.~E.~Ibanez,
  ``Modulus-dominated SUSY-breaking soft terms in F-theory and their test at
  LHC,''
  JHEP {\bf 0807} (2008) 099
  [arXiv:0805.2943 [hep-ph]].

\bibitem{Marsano:2008jq}
  J.~Marsano, N.~Saulina and S.~Schafer-Nameki,
  ``Gauge Mediation in F-Theory GUT Models,''
  Phys.\ Rev.\  D {\bf 80} (2009) 046006
  [arXiv:0808.1571 [hep-th]].

\bibitem{Heckman:2008qt}
  J.~J.~Heckman and C.~Vafa,
  ``F-theory, GUTs, and the Weak Scale,''
  arXiv:0809.1098 [hep-th].
  
\bibitem{Font:2008id}
  A.~Font and L.~E.~Ibanez,
  ``Yukawa Structure from U(1) Fluxes in F-theory Grand Unification,''
  JHEP {\bf 0902} (2009) 016
  [arXiv:0811.2157 [hep-th]].

\bibitem{Heckman:2008qa}
  J.~J.~Heckman and C.~Vafa,
  ``Flavor Hierarchy From F-theory,''
  arXiv:0811.2417 [hep-th].

\bibitem{Blumenhagen:2008aw}
  R.~Blumenhagen,
  ``Gauge Coupling Unification In F-Theory Grand Unified Theories,''
  Phys.\ Rev.\ Lett.\  {\bf 102} (2009) 071601
  [arXiv:0812.0248 [hep-th]].

\bibitem{09013785}
  J.~L.~Bourjaily,
  ``Local Models in F-Theory and M-Theory with Three Generations,''
  arXiv:0901.3785 [hep-th].

\bibitem{Hayashi:2009ge}
  H.~Hayashi, T.~Kawano, R.~Tatar and T.~Watari,
  ``Codimension-3 Singularities and Yukawa Couplings in F-theory,''
  Nucl.\ Phys.\  B {\bf 823} (2009) 47
  [arXiv:0901.4941 [hep-th]].

  \bibitem{09033009}
  C.~M.~Chen and Y.~C.~Chung,
  ``A Note on Local GUT Models in F-Theory,''
  Nucl.\ Phys.\  B {\bf 824} (2010) 273
  [arXiv:0903.3009 [hep-th]].

\bibitem{Donagi:2009ra}
  R.~Donagi and M.~Wijnholt,
  ``Higgs Bundles and UV Completion in F-Theory,''
  arXiv:0904.1218 [hep-th].

\bibitem{Bouchard:2009bu}
  V.~Bouchard, J.~J.~Heckman, J.~Seo and C.~Vafa,
  ``F-theory and Neutrinos: Kaluza-Klein Dilution of Flavor Hierarchy,''
  arXiv:0904.1419 [hep-ph].

\bibitem{Randall:2009dw}
  L.~Randall and D.~Simmons-Duffin,
  ``Quark and Lepton Flavor Physics from F-Theory,''
  arXiv:0904.1584 [hep-ph].

  \bibitem{09043101}
  J.~J.~Heckman and C.~Vafa,
  ``CP Violation and F-theory GUTs,''
  arXiv:0904.3101 [hep-th].

\bibitem{09043932}
  J.~Marsano, N.~Saulina and S.~Schafer-Nameki,
  ``F-theory Compactifications for Supersymmetric GUTs,''
  JHEP {\bf 0908} (2009) 030
  [arXiv:0904.3932 [hep-th]].

\bibitem{Marsano:2009wr}
  J.~Marsano, N.~Saulina and S.~Schafer-Nameki,
  ``Compact F-theory GUTs with $U(1)_{PQ}$,''
  arXiv:0912.0272 [hep-th].

\bibitem{09052289}
  R.~Tatar, Y.~Tsuchiya and T.~Watari,
  ``Right-handed Neutrinos in F-theory Compactifications,''
  Nucl.\ Phys.\  B {\bf 823} (2009) 1
  [arXiv:0905.2289 [hep-th]].

\bibitem{Heckman:2009mn}
  J.~J.~Heckman, A.~Tavanfar and C.~Vafa,
  ``The Point of E8 in F-theory GUTs,''
  arXiv:0906.0581 [hep-th].

\bibitem{Marsano:2009gv}
J.~Marsano, N.~Saulina and S.~Schafer-Nameki,
``Monodromies, Fluxes, and Compact Three-Generation F-theory GUTs,''
arXiv:0906.4672 [hep-th].

\bibitem{09063297}
  R.~Blumenhagen, J.~P.~Conlon, S.~Krippendorf, S.~Moster and F.~Quevedo,
  ``SUSY Breaking in Local String/F-Theory Models,''
  JHEP {\bf 0909} (2009) 007
  [arXiv:0906.3297 [hep-th]].

\bibitem{Conlon:2009qa}
  J.~P.~Conlon and E.~Palti,
  ``On Gauge Threshold Corrections for Local IIB/F-theory GUTs,''
  arXiv:0907.1362 [hep-th].

\bibitem{Font:2009gq}
  A.~Font and L.~E.~Ibanez,
  ``Matter wave functions and Yukawa couplings in F-theory Grand Unification,''
  arXiv:0907.4895 [hep-th].

\bibitem{08070789}
  J.~P.~Conlon, A.~Maharana and F.~Quevedo,
  ``Wave Functions and Yukawa Couplings in Local String Compactifications,''
  JHEP {\bf 0809} (2008) 104
  [arXiv:0807.0789 [hep-th]].

\bibitem{Cecotti:2009zf}
  S.~Cecotti, M.~C.~N.~Cheng, J.~J.~Heckman and C.~Vafa,
  ``Yukawa Couplings in F-theory and Non-Commutative Geometry,''
  arXiv:0910.0477 [hep-th].

\bibitem{Conlon:2009qq}
  J.~P.~Conlon and E.~Palti,
  ``Aspects of Flavour and Supersymmetry in F-theory GUTs,''
  arXiv:0910.2413 [hep-th].

\bibitem{Hayashi:2009bt}
  H.~Hayashi, T.~Kawano, Y.~Tsuchiya and T.~Watari,
  ``Flavor Structure in F-theory Compactifications,''
  arXiv:0910.2762 [hep-th].

\bibitem{Marchesano:2009rz}
  F.~Marchesano and L.~Martucci,
  ``Non-perturbative effects on seven-brane Yukawa couplings,''
  arXiv:0910.5496 [hep-th].

\bibitem{Chung:2009ib}
  Y.~C.~Chung,
  ``Abelian Gauge Fluxes and Local Models in F-Theory,''
  arXiv:0911.0427 [hep-th].

\bibitem{Blumenhagen:2008zz}
  R.~Blumenhagen, V.~Braun, T.~W.~Grimm and T.~Weigand,
  ``GUTs in Type IIB Orientifold Compactifications,''
  Nucl.\ Phys.\  B {\bf 815} (2009) 1
  [arXiv:0811.2936 [hep-th]].

\bibitem{09024143}
  B.~Andreas and G.~Curio,
  ``From Local to Global in F-Theory Model Building,''
  arXiv:0902.4143 [hep-th].

\bibitem{09060013}
  R.~Blumenhagen, T.~W.~Grimm, B.~Jurke and T.~Weigand,
  ``F-theory uplifts and GUTs,''
  JHEP {\bf 0909} (2009) 053
  [arXiv:0906.0013 [hep-th]].

\bibitem{Blumenhagen:2009yv}
  R.~Blumenhagen, T.~W.~Grimm, B.~Jurke and T.~Weigand,
  ``Global F-theory GUTs,''
  arXiv:0908.1784 [hep-th].

\bibitem{Katz:1996xe}
  S.~H.~Katz and C.~Vafa,
  ``Matter from geometry,''
  Nucl.\ Phys.\  B {\bf 497}, 146 (1997)
  [arXiv:hep-th/9606086].

\bibitem{Bershadsky:1996nu}
  M.~Bershadsky and A.~Johansen,
  ``Colliding singularities in F-theory and phase transitions,''
  Nucl.\ Phys.\  B {\bf 489}, 122 (1997)
  [arXiv:hep-th/9610111].

\bibitem{Witten:1995gx}
  E.~Witten,
  ``Small Instantons in String Theory,''
  Nucl.\ Phys.\  B {\bf 460}, 541 (1996)
  [arXiv:hep-th/9511030].

\bibitem{Froggatt:1978nt}
  C.~D.~Froggatt and H.~B.~Nielsen,
  ``Hierarchy Of Quark Masses, Cabibbo Angles And CP Violation,''
  Nucl.\ Phys.\  B {\bf 147}, 277 (1979).

\bibitem{Dreiner:2003hw}
  H.~K.~Dreiner and M.~Thormeier,
  ``Supersymmetric Froggatt-Nielsen models with baryon- and lepton-number
  violation,''
  Phys.\ Rev.\  D {\bf 69} (2004) 053002
  [arXiv:hep-ph/0305270].

\bibitem{Babu:2009fd}
  K.~S.~Babu,
  ``TASI Lectures on Flavor Physics,''
  arXiv:0910.2948 [hep-ph].

\bibitem{Conlon:2007zza}
  J.~P.~Conlon and D.~Cremades,
  ``The neutrino suppression scale from large volumes,''
  Phys.\ Rev.\ Lett.\  {\bf 99} (2007) 041803
  [arXiv:hep-ph/0611144].

\bibitem{Conlon:2009xf}
  J.~P.~Conlon,
  ``Gauge Threshold Corrections for Local String Models,''
  JHEP {\bf 0904} (2009) 059
  [arXiv:0901.4350 [hep-th]].

\bibitem{Conlon:2009kt}
  J.~P.~Conlon and E.~Palti,
  ``Gauge Threshold Corrections for Local Orientifolds,''
  JHEP {\bf 0909} (2009) 019
  [arXiv:0906.1920 [hep-th]].

\bibitem{Barbier:2004ez}
  R.~Barbier {\it et al.},
  ``R-parity violating supersymmetry,''
  Phys.\ Rept.\  {\bf 420} (2005) 1
  [arXiv:hep-ph/0406039].

\bibitem{Smirnov:1996bg}
  A.~Y.~Smirnov and F.~Vissani,
  ``Upper bound on all products of R-parity violating couplings $\lambda'$ and
  $\lambda''$ from proton decay,''
  Phys.\ Lett.\  B {\bf 380} (1996) 317
  [arXiv:hep-ph/9601387].

\bibitem{Giudice:1988yz}
  G.~F.~Giudice and A.~Masiero,
  ``A Natural Solution to the mu Problem in Supergravity Theories,''
  Phys.\ Lett.\  B {\bf 206} (1988) 480.

\bibitem{Strumia:2006db}
  A.~Strumia and F.~Vissani,
  ``Neutrino masses and mixings and.,''
  arXiv:hep-ph/0606054.

\bibitem{Amsler}
C.~Amsler~et~al. (Particle Data Group), Physics Letters B667, 1 (2008) 

\bibitem{ArkaniHamed:2000bq}
  N.~Arkani-Hamed, L.~J.~Hall, H.~Murayama, D.~Tucker-Smith and N.~Weiner,
  ``Small neutrino masses from supersymmetry breaking,''
  Phys.\ Rev.\  D {\bf 64} (2001) 115011
  [arXiv:hep-ph/0006312].

\bibitem{Hall:1993ni}
  L.~J.~Hall and A.~Rasin,
  ``On The Generality Of Certain Predictions For Quark Mixing,''
  Phys.\ Lett.\  B {\bf 315} (1993) 164
  [arXiv:hep-ph/9303303].

E.~Dudas, S.~Pokorski and C.~A.~Savoy,
``Yukawa matrices from a spontaneously broken Abelian symmetry,''
Phys.\ Lett.\  B {\bf 356} (1995) 45
arXiv:hep-ph/9504292].


\bibitem{Buican:2006sn}
  M.~Buican, D.~Malyshev, D.~R.~Morrison, H.~Verlinde and M.~Wijnholt,
  ``D-branes at singularities, compactification, and hypercharge,''
  JHEP {\bf 0701} (2007) 107
  [arXiv:hep-th/0610007].


\end{thebibliography}
\end{document}